\documentclass[10pt]{article}
\usepackage[tbtags]{amsmath}
\usepackage[linktocpage]{hyperref}
\hypersetup{colorlinks, linkcolor={red!60!black}, citecolor={blue!60!black}, urlcolor={blue!80!black}}
\usepackage{amssymb,amsthm,mathrsfs}
\usepackage{amsfonts}
\usepackage{graphicx}
\usepackage{stackrel}
\usepackage{subcaption}
\graphicspath{{figs/}}
\usepackage[left=2.5cm,right=2.2cm,top=0.5cm,bottom=1.3cm,includeheadfoot]{geometry}
\usepackage{indentfirst}
\usepackage{color}
\usepackage{xcolor}
\usepackage{cite}
\usepackage{mathtools}
\usepackage{tikz-cd}
\usepackage{enumitem}
\usepackage{multicol}
\usepackage{float}
\usepackage[normalem]{ulem}
\usepackage{caption}
\usepackage{subcaption}
\definecolor{PV}{RGB}{120,81,169}

\usepackage{svg}

\numberwithin{equation}{section}

\renewcommand*{\thefootnote}{\fnsymbol{footnote}}

\begin{document}
	\begin{center}
		{\Large\bf
		Hawking radiation and graybody factor of slightly deformed Schwarzschild black holes
        }\\
		\vskip 5mm
		{\large
			Asier Alonso-Bardaji\footnote{e-mail address: {\tt asier.alonso@ehu.eus}},
			David Brizuela
			\footnote{e-mail address: {\tt david.brizuela@ehu.eus}},
			and Marc Schneider
			\footnote{e-mail address: {\tt marc.schneider@ehu.eus}}}
		\vskip 3mm
		{\sl 
		Department of Physics and EHU Quantum Center, University of the Basque Country,\\
			Barrio Sarriena s/n, 48940 Leioa, Spain}\\\vskip 1mm
	\end{center}

\setcounter{footnote}{0}
\renewcommand*{\thefootnote}{\arabic{footnote}}

	\begin{abstract}
We analyze the radiative properties of a general class of static and spherically symmetric black holes described by three independent shape functions,
that encompass a broad family of regular and quantum-corrected geometries.
We consider a minimally coupled massless scalar field, and derive the Hawking temperature and graybody factor for the different geometries.
While the Hawking temperature is determined by the near-horizon geometry, the graybody factor
depends on the scattering properties of the entire exterior spacetime, and therefore exhibits a richer dependence on the deformation parameters, frequency, and angular mode.
Then we particularize to geometries that describe slight deformations of the Schwarzschild black hole, and obtain analytic expressions
for the leading corrections to the graybody factor. Finally, we apply our results to several specific well-known black-hole models,
including Reissner-Nordström, Bardeen, Hayward, Simpson-Visser, as well as geometries inspired by loop quantum gravity.
In general,
the deformations lead to colder horizons than Schwarzschild, and their effect on the graybody factor
is frequency-dependent: the transmission of low-frequency modes is suppressed, whereas it is enhanced for high-frequency modes.
This behavior is found for all but one of the geometries considered, which exhibits just the opposite trend. 
The analyzed modifications to the Hawking spectrum may therefore have relevant observational consequences,
potentially providing signatures of deviations from the Schwarzschild geometry.
\end{abstract}

\section{Introduction}\label{sec:intro}

Black holes are one of the most remarkable predictions of general relativity. However, a significant challenge in the general understanding of these objects hides deep behind their horizons. There, one classically encounters the singularity, suggesting the need for modifications of the theory at high-curvature regimes. The occurrence of these singularities has motivated the construction of various regular black-hole solutions by strategically modifying the assumptions in singularity theorems, including modifications of the gravitational dynamics, addition of exotic matter, or effective descriptions of short-distance gravitational physics.
Modified black holes were first introduced in the context of specific models such as the Bardeen geometry~\cite{bardeen}, and later generalized to a much broader family of models, including the Hayward~\cite{Hayward:2005gi} and Simpson-Visser~\cite{Simpson:2018tsi} spacetimes. These models are generically built to
asymptotically recover the Schwarzschild geometry at large distances outside the horizon, leaving their differences mainly confined to the region that is causally disconnected from the exterior. Whether such effects can be observed from asymptotic regions remains an open question.

Hawking radiation provides a particularly interesting framework to address this issue. As it is well known, Hawking radiation covers the thermal emission of black holes,
establishing a connection between horizon physics and
asymptotic observers~\cite{hawking1974,hawking1975}. For stationary black holes, the observed spectrum depends on two key ingredients:
the horizon temperature, which is a local property controlled by the near-horizon geometry, and the graybody factor, which is the frequency-dependent transmission coefficient for the outgoing modes in the scattering off the black hole's potential barrier.
Modulations in any of these quantities may encode information about deviations from general relativity.

There is a substantial literature about Hawking radiation in modified and nonsingular black holes~\cite{Rincon:2020cos,Arbey:2021jif,Arbey:2021yke,calza2025grayhawk,calza2025primordial1,calza2025primordial2}. In particular, the graybody factor has been computed for several classes of regular geometries,
such as the Bardeen~\cite{Konoplya:2023ahd,Lutfuoglu:2026rqe}, Hayward~\cite{Konoplya:2023ppx},
and Simpson-Visser~\cite{Jha:2023wzo} black holes. Some recent proposals motivated by loop quantum gravity \cite{Alonso-Bardaji:2021yls,Alonso-Bardaji:2022ear,Belfaqih:2024vfk} also show interesting deviations from the classical emission spectrum~\cite{Alonso-Bardaji:2025qft,Belfaqih:2026qgj}. All these results emphasize that the propagation problem, which clearly depends on the particular geometry, can carry information about the underlying high-curvature regime.

Despite the growing literature, most studies focus on one particular model. Due to the inherent differences between each spacetime, this makes it difficult to determine generic and particular predictions from modifying the Schwarzschild geometry. The casuistry of the modified spectra is also large, and the corrections to the temperature and to the scattering problem do not need to be correlated. This distinction was illustrated explicitly in a recent analysis of a nonsingular black-hole geometry with a minimal surface replacing the classical singularity~\cite{Alonso-Bardaji:2025qft}. The resulting spectrum turned out to be colder and purer.

In the present paper, we address this problem by considering a general parameterization of static spherical black holes, which includes as particular cases the above mentioned regular geometries. The parameterization employs three shape functions that arise naturally from the Hamiltonian description of spherical gravity developed in Ref.~\cite{Alonso-Bardaji:2025hda}. This formalism allows matter fields to be dynamically coupled to the geometry in a straightforward way, providing a natural method to study modified gravitational dynamics. Nevertheless, here we take a more conservative approach and treat the geometry as a fixed background over which a minimally coupled test scalar field propagates. In this way, we isolate the gravitational effects on the emitted radiation, allowing us to identify the features of the geometric corrections.

Specifically, we will see how different shape functions in the metric influence both local (temperature) and global (graybody factor) properties of the radiation spectrum in qualitatively different ways. It is important to remark that, in order to compare different geometries, we need to agree on some common feature of the black-hole spacetimes. Here, we provide a systematic way to compare the radiative properties of black holes of equal size, that is, with the same horizon area. This methodology then permits to contrast the radiative properties of different models based on a geometrically well-defined reference scale that is directly observable.
In particular,
we will focus on small deviations from the Schwarzschild geometry, as probes for physics beyond general relativity. 
For this class of geometries, as it will be shown below, we can perturbatively expand the modifications and obtain a four-parameter family of geometries,
whose specific effects on the horizon temperature and graybody factor can be identified separately.
In this way, we provide a framework for comparing the radiative properties of different black-hole models and quantifying their deviations from the Schwarzschild geometry.

Our analysis is organized as follows: in Sec.~\ref{sec:geometry} we introduce a family of general, static, and spherically symmetric
geometries parameterized in terms of three shape functions of the
area radius, and discuss their main geometric properties. 
In Sec.~\ref{sec.scalar}, we study a massless scalar field propagating on these backgrounds. Sec.~\ref{sec.radiation} provides a general derivation of the Hawking temperature and graybody factors. In Sec.~\ref{sec_deformedschwarzschild} we thoroughly study the radiative properties of slightly modified Schwarzschild black holes.  As an application, Sec.~\ref{sec:apltpbhsts} shows the specific results for some well-known spherically symmetric, static 
black-hole geometries. This last section is written in an almost self-contained
form; the reader interested in these particular results can proceed directly to this section. We end the main body of the paper with some concluding remarks in Sec.~\ref{sec.concl}.

\section{Geometry}
\label{sec:geometry}

We begin our analysis by assuming a general spherically symmetric
four dimensional manifold given by the warped product ${\cal M}_2\times \mathbb{S}_2$,
with ${\cal M}_2$ being a two-dimensional Lorentzian manifold and $\mathbb{S}_2$ the two-sphere. Any metric on this manifold can be written as
\begin{equation}\label{eq:warped}
    {}^{(4)}g=g+r^2\mbox{d}\mathbb{S}_2, 
\end{equation}
where $\mbox{d}\mathbb{S}_2=\mbox{d}\vartheta\otimes\mbox{d}\vartheta+\sin^2(\vartheta)\mbox{d}\varphi\otimes\mbox{d}\varphi$
is the standard metric on the two-sphere and $g$ a two-dimensional metric on ${\cal M}_2$. The areal radius $r$ is a positive-definite scalar field on $\mathcal{M}_2$, such that the associated spherical orbits
have an area of $4 \pi r^2$.

Our goal is to study the properties of a scalar field propagating
on the exterior of a spherical black-hole geometry.
Therefore, we choose a diagonal chart, with the scalar $r$ describing the radial coordinate,
and introduce the following parameterization,
\begin{equation}\label{eq:metrikh123}
       {}^{(4)}g=-\left(h_1(r)-\frac{2M}{h_2(r)}\right)\mbox{d}t\otimes\mbox{d}t+\frac{\mbox{d}r\otimes\mbox{d}r}{h_3(r)\left(h_1(r)-\frac{2M}{h_2(r)}\right)}+r^2\mbox{d}\mathbb{S}_2,
\end{equation}
in terms of the three shape functions $h_1(r)$, $h_2(r)$, and $h_3(r)$, and the positive constant $M>0$. The reference values of these shape functions $h_1(r)=1=h_3(r)$ and $h_2(r)=r$ correspond to the Schwarzschild geometry. 
As will be explicitly shown below, this parameterization is remarkably general and it describes a broad class of
spherically symmetric static spacetimes, such as Reissner-Nordstr\"om, Bardeen, and Hayward black holes,
as well as several recent proposals in the context of quantum gravity.
In addition, a compelling motivation for this parameterization comes from Ref.~\cite{Alonso-Bardaji:2025hda},
where this family of metrics was equipped with a simple Hamiltonian formulation.
Such formulation provides a well-defined and systematic framework
for coupling matter fields to the corresponding vacuum geometry in order to
explore collapsing models and matter backreaction on the geometry.
However, in the present paper we will not be concerned about the dynamical
origin of the metric \eqref{eq:metrikh123}, and it will be treated as a given
fixed background on which test matter fields propagate.

By construction, $\partial_t$ is a Killing vector of the geometry \eqref{eq:metrikh123}.
Its squared norm is given
by
\begin{equation}
G(r):=-\left(h_1(r)-\frac{2M}{h_2(r)}\right),
\end{equation}
and it allows us to re-express the metric
in a compact form as
\begin{equation}\label{eq:metrikG-param}
      {}^{(4)} g=G(r)\mbox{d}t\otimes\mbox{d}t-\frac{\mbox{d}r\otimes\mbox{d}r}{G(r)h_3(r)}+r^2\mbox{d}\mathbb{S}_2.
\end{equation}
The vanishing of $G(r_K)=0$, which implies $h_1(r_K)h_2(r_K)=2M$,
marks the position of a Killing horizon at $r=r_K$ and also the limit of the domain of definition of this chart.
Of course, for generic functions $h_1(r)$ and $h_2(r)$,
such condition can be reached by various configurations of the product $h_1 h_2$, leading to a myriad of horizons.

However,
since we are interested in studying deformations of the Schwarzschild geometry, we rely on three specific assumptions, which impose a number of
restrictions on the shape functions.

\begin{itemize}
\item First, the spacetime must be a Lorentzian manifold, and thus $h_3>0$ in the domain of validity of the chart \eqref{eq:metrikG-param}.
\item Second, we require the existence of
a nondegenerate outermost horizon at $r=r_H$ with an external static and asymptotically flat region.
That is, $r_H$ is defined
to be a simple root of $G(r_H)=0$, with $G'(r_H)< 0$, such that, for all $r>r_H$,
the Killing field is timelike, i.e., $G(r)<0$, and thus $h_1(r)h_2(r)> 2 M$.
Asymptotic flatness is guaranteed by assuming that the shape functions tend to
\begin{align}\label{eq.decays}
h_1(r) &=1+{\cal O}(1/r^2)\nonumber,\\
h_2(r)&=r+ {\cal O}(1),\\
h_3(r)&=1+ {\cal O}(1/r)\nonumber,
\end{align}
for $r\to \infty$.
Note, in particular, that we are assuming that $h_1$ tends to 1 as $1/r^2$,
which is faster than strictly necessary for asymptotic flatness.
Nonetheless, this does not imply any restriction,
as asymptotically any term proportional to $1/r$
in $h_1$ can simply be absorbed in a redefinition of the parameter $M$.
In this way, the squared norm of the Killing field
asymptotically tends to the Schwarzschild form $G(r)=-1+2M/r+{\cal O}(1/r^2)$,
and there are no parameters other than $M$ appearing at that order.
However, as will be explained below, the present
assumptions still allow certain standard mass functions
to take values different from $M$.
\item Third, the region of the spacetime under consideration should be smooth, and thus the three shape functions $h_1$, $h_2$, and $h_3$ are regular and finite for all $r\geq r_H$.
\end{itemize}

Finally, we note that all these assumptions impose conditions on the shape functions
in the region $r\geq r_H$. Nonetheless, we are not imposing anything, and the
shape functions are thus completely free, for the interior region $r<r_H$.
Therefore, all the results of the present paper will be insensitive to the internal
structure of the black hole, and, even if we are specially interested in
singularity-free black holes, all results apply also to a wide variety of black-hole geometries with different
internal structures, such as multiple horizons
or wormholes.

\subsection{Curvature}

The curvature of these black holes is completely determined in terms of the four-dimensional Ricci scalar,
\begin{align}\label{ricci}
   {}^{(4)}R&=R+\frac{2}{r^2}\left(1-g(\nabla r, \nabla r)\right)-\frac{4}{r}\,\Box r,
\end{align}
and the Weyl scalar
\begin{align}\label{psi2}
\Psi_2 &={-\frac{1}{12}}\left(R+\frac{2}{r^2}(1-g(\nabla r, \nabla r))+\frac{2}{r}\,\Box r\right),
\end{align}
where $R$ is the curvature and $\Box$ the Laplace-Beltrami operator corresponding to the two-dimensional metric $g$ on ${\cal M}_2$.
In terms of the shape functions, these quantities read explicitly
\begin{align}
      {}^{(4)} R=&-\frac{1}{2 r^2 h_2^3}[h_2^3 (r^2 h_1' h_3'+2 r h_3
   (r h_1''+4 h_1')+4 h_1(r
   h_3'+h_3)-4)+8 M r^2 h_3
   h_2'^2\nonumber\\
   &-2 M r h_2(r h_2' h_3'+2 h_3
   (r h_2''+4 h_2'))+8 M h_2^2 (r
   h_3'+h_3)],
\end{align}
\begin{align}
  \Psi_2=&\;  \frac{1}{24 r^2 h_2^3}[h_2^3 (r (2 r h_3 h_1''+h_1'(r
   h_3'-4 h_3))+h_1(4 h_3-2 r
   h_3')-4)-8 M r^2 h_3 h_2'^2\nonumber\\
   &+2 M r h_2
   (2 r h_3 h_2''+h_2'(r h_3'-4
   h_3))+4 M h_2^2 (r h_3'-2
   h_3)],
\end{align}
where a prime denotes a derivative with respect to $r$.

\subsection{The concept of mass}\label{sec.mass}

In the context of metric theories of gravity, there exist several definitions
of mass. Depending on the geometry, such concepts of mass may or may not coincide. In this section we review three different standard definitions
of mass and evaluate them for the geometry described by the
line element \eqref{eq:metrikG-param}.

\subsubsection{Misner-Sharp mass}

The Misner-Sharp mass is
defined quasi-locally in terms of
the gradient of the areal radius function:
\begin{align}
    M_{\rm MS}:=\frac{r}{2}\left(1-g(\nabla r, \nabla r)\right)=\frac{r}{2}\left(1+h_3G\right).
\end{align}
Note that this is proportional to the second term in \eqref{ricci} and \eqref{psi2}, and thus encodes part of the curvature of the spacetime.
The Misner-Sharp mass is supposed to encode the gravitating energy
contained inside a finite sphere of radius $r$, but
there is no guaranteed monotonicity increase of $M_{\rm MS}$
with $r$ unless $1+h_3 G>-r(h_3 G)'$. Nonetheless,
since $G$ vanishes at Killing horizons $r=r_K$, there is a proportional relation
between $M_{\rm MS}(r_K)$ and $r_K$ there,
that is, $r_K=2 M_{\rm MS}(r_K)$. And, for the multihorizon case,
the value of $M_{\rm MS}$ evaluated at
consecutive Killing horizons would indeed be linearly increasing with $r$.

For the Schwarzschild
geometry $h_3G=-1+2 M/r$, and thus $M_{\rm MS}$ is constant and equal to the
metric parameter $M_{\rm MS}=M$. However, we note that this is a very
special property of such geometry, while in general $M_{\rm MS}$ depends
on $r$. For instance, taken into account the asymptotic decays
\eqref{eq.decays} assumed above,
the Misner-Sharp mass tends to
\begin{equation}\label{eq.MSinf}
\lim_{r\to\infty} M_{\rm MS}=M+\frac{1}{2}\lim_{r \to\infty}r^2 h_3'(r).
\end{equation}
Therefore, at $r\to\infty$ we get $M_{\rm MS}=M$ only if $h_3$ tends to 1 as $1/r^2$ or faster.
Otherwise, the order $1/r$ in $h_3(r)$ contributes to the asymptotic value of $M_{\rm MS}$,
which differs from $M$ \cite{Alonso-Bardaji:2022ear}.

\subsubsection{Arnowitt-Deser-Misner mass}

The Arnowitt-Deser-Misner mass $M_{\rm ADM}$ is defined at the asymptotic flat spatial infinity as
the integral over a closed surface $\mathcal{S}$ with outward normal $s^i$ \cite{Szabados:2009eka},
\begin{align}
    M_{\rm ADM} := \frac{1}{16\pi}\lim_{r\to\infty}\oint_\mathcal{S} \sum_{i,j}\left(\frac{\partial q_{ij}}{\partial x^j}-\frac{\partial q_{jj}}{\partial x^i}\right) s^i {\rm d}\mathcal{S}.
\end{align}
Here $q_{ij}$, with $i=1,2,3$, are the components of the metric of the three-dimensional hypersurfaces of constant $t$
in terms of the Cartesian coordinates $x^i$.
If one assumes that $\mathcal{S}$ is a sphere of constant radius $r$, this expression reduces to
\begin{align}
    M_{\rm ADM} = \lim_{r\to\infty} r\left(1-\sqrt{-h_3G}\right).
\end{align}
Therefore, the value of $M_{\rm ADM}$ depends on the asymptotic fall-off of the function $\sqrt{-h_3G}$.
More precisely,
following the decay conditions \eqref{eq.decays} we have imposed on the shape functions, we can write
\begin{align}
 M_{\rm ADM} =M+\frac{1}{2}\lim_{r\to\infty} r^2 h_3'(r),
 \end{align}
 which has the same form as \eqref{eq.MSinf}.
Therefore, the ADM mass of the constant $t$-slices coincides with the asymptotic value
of the Misner-Sharp mass \cite{https://doi.org/10.1002/cpa.3160390505},
\begin{equation}
M_{\rm ADM}=\lim_{r\to\infty}M_{MS}.
\end{equation}

\subsubsection{Komar mass}

The Komar mass provides yet a different notion of mass for stationary spacetimes and it is defined
by the integral
\begin{align}
    M_{\rm K}:=\frac{1}{8\pi}\int_{\partial\mathcal{S}}\star{\rm d}\chi,
\end{align}
over the boundary of the closed surface $\mathcal{S}$,
where $\chi$ is the timelike Killing field and $\star$ the Hodge-$\star$ operator.
If the surface $\mathcal{S}$ is chosen to be
a sphere of constant $t$ and $r$, we get
\begin{align}
     M_{\rm K}=-\frac{r^2}{2}\sqrt{h_3}G'(r).
\end{align}
As the Misner-Sharp mass, $M_{\rm K}$ is a quasi-local function that depends on $r$
but it does not have a clear local interpretation in terms of gravitating energy. 
However, the value of $M_{\rm K}$
at the Killing horizon is proportional to the surface gravity, and thus it encodes the temperature of the horizon. 
Concerning its behavior at spatial infinity, one gets the constant parameter $M$, i.e.,
\begin{equation}
\lim_{r\to\infty}M_{\rm K}=M,
\end{equation}
for the decay of the shape functions \eqref{eq.decays}.

\subsubsection{Summary}

To sum up, we have introduced two quasi-local, $M_{\rm MS}$ and $M_{\rm K}$, and a global $M_{\rm ADM}$ mass functions.
In general, at any given radius $r$, the values $M_{\rm MS}(r)$ and $M_{\rm K}(r)$ will not coincide,
although they are equivalent in general relativity, i.e., for any geometry satisfying Einstein's
field equations $M_{\rm MS}(r)=M_{\rm K}(r)$ \cite{BEIG1978153}. 
In the asymptotic limit $r\to\infty$ it turns out that in general $M_{\rm MS}\to M_{\rm ADM}$, while $M_{\rm K}\to M$. Only if the function $h_3$
asymptotes as $h_3(r)=1+{\cal O}(1/r^2)$, all the three mass functions asymptotically tend to the constant parameter $M$.

All this discussion shows that it is not straightforward to define the mass of a given black hole, let alone the gravitating energy of different black-hole families. Since our goal is to compare the radiative properties of slightly deformed Schwarzschild black holes,
it is necessary to agree on a common magnitude for the different geometries, and contrast, in this way,
the effects of the different realizations of the shape functions $h_1$, $h_2$, and $h_3$ in \eqref{eq:metrikh123}. 
This is the reason why in the analysis presented in Secs.~\ref{sec_deformedschwarzschild} and \ref{sec:apltpbhsts} we will opt to
compare black holes of ``equal size'', that is, those with the same value of the horizon area $4 \pi r_H^2$.
In particular, as explained above, this also implies that each of these black holes will have the same Misner-Sharp
mass evaluated on the horizon $M_{\rm MS}(r_H)=\frac{1}{2}r_H$, but
its asymptotic limit for different geometries, and thus their corresponding $M_{\rm ADM}$,  may differ.

\section{A scalar test field: the Regge-Wheeler potential}\label{sec.scalar}

To analyze the radiative properties of this family of geometries, we will consider
a massless, minimally coupled, scalar test field $\phi$, which obeys the Klein-Gordon equation ${}^{(4)}\Box\phi=0$.
Making use of the warped form of the metric \eqref{eq:warped},
the four dimensional d'Alembert operator ${}^{(4)}\Box$ can be decomposed as 
\begin{equation}\label{eq:KGallg}
 {}^{(4)}\Box\phi=\Box\phi +\frac{2}{r} g(\nabla r,\nabla \phi) + \frac{1}{r^2}\Delta_\sphericalangle\phi,
\end{equation}
with $\Box$ and $\Delta_\sphericalangle$ being the Laplace-Beltrami operator on ${\cal M}_2$ and $\mathbb{S}_2$, respectively. 
Due to the symmetry of the background geometry, it is natural to perform a decomposition of the field into spherical harmonics
\begin{equation}
\phi(x)= \frac{1}{r}\sum_{\ell=0}^\infty \sum_{m=-\ell}^{\ell}\,\psi_{\ell}(t,r)\,Y_{\ell m} (\vartheta,\varphi),
\end{equation}
where a global $1/r$ factor is introduced for convenience.
Note also that, since the dynamics of the modes does not depend on the magnetic number $m$,
we omit it as an additional subindex in the mode functions, and simply refer to them as $\psi_\ell$.
By introducing this decomposition in the Klein-Gordon equation \eqref{eq:KGallg},
and using that $\Delta_\sphericalangle Y_{\ell m} (\vartheta,\varphi)=-\ell(\ell+1)Y_{\ell m} (\vartheta,\varphi)$,
one gets the following partial differential equation for the modes on $\mathcal{M}_2$:
\begin{equation}\label{eq_modeequation}
\Box\psi_{\ell}-\left(\frac{\Box r}{r}+\frac{\ell(\ell+1)}{r^2}\right)\psi_{\ell}=0.
\end{equation}
Now, we take into account the explicit coordinate representation of the d'Alembert operator for any diagonal static metric on $\mathcal{M}_2$ as given
in \eqref{eq:metrikG-param},
\begin{equation}\label{eq.boxcoords}
\Box\psi_{\ell}=\frac{1}{G} \partial^2_t\psi_l - \sqrt{h_3} \partial_r (\sqrt{h_3}G\partial_r\psi_l).
\end{equation}
We define the tortoise coordinate $r_*$ through the differential
\begin{equation}\label{eq.tortoise}
   \frac{\mbox{d}r_\ast}{\mbox{d}r}=\frac{-1}{G(r)\sqrt{h_3(r)}}=:\frac{1}{n(r)},
\end{equation}
such that the two-dimensional metric takes a conformally flat form
\begin{equation}\label{flatmetric}
g=G(r_*)(\mbox{d}t\otimes\mbox{d}t-\mbox{d}r_*\otimes\mbox{d}r_*),
\end{equation}
and the term in parentheses in \eqref{eq.boxcoords} becomes a total derivative, that is,
$\sqrt{h_3} G \partial_r\psi_\ell=-\partial_{r_*}\psi_\ell$.
In this way, in terms of $r_*$, the action of the box operator simplifies,
\begin{equation}\label{eq.boxtortoise}
\Box\psi_{\ell}(t,r_\ast)=\frac{1}{G(r_\ast)}\left(\partial_t^2\psi_{\ell}(t,r_\ast)-\partial_{r_*}^2\psi_{\ell}(t,r_\ast)\right).
\end{equation}
With this form, the mode equation \eqref{eq_modeequation} can be rewritten as
\begin{equation}\label{eq:RWG}
-\frac{\partial^2\psi_{\ell}(t,r_\ast)}{\partial t^2}+\frac{\partial^2\psi_{\ell}(t,r_\ast)}{\partial r_*^2}- V_\ell(r(r_\ast)) \psi_{\ell}(t,r_\ast)=0,
\end{equation}
where we have defined the Regge-Wheeler potential,
\begin{equation}\label{potential}
V_\ell(r):=-{G}(r)\,\left(\frac{\ell(\ell+1)}{r^2}+\frac{\Box r}{r}\right).
\end{equation}
Therefore, in terms of the tortoise coordinate $r_*$, the dynamics is described by
the wave equation \eqref{eq:RWG} on a two-dimensional Minkowski background, while all the curvature effects are
encoded in the potential \eqref{potential}. This potential is composed by two terms.

On the one hand, the centrifugal term
$-G(r)\frac{\ell(\ell+1)}{r^2}$, which appears rescaled by the squared norm of the Killing $G$,
originates from the curvature of the sphere and it depends on the mode number $\ell$. Since $G<0$
in the region under consideration, the centrifugal term is positive
definite for all the domain $r\in(r_H,\infty)$. It is vanishing at the horizon $r=r_H$
and, in the limit $r\to\infty$, it decays to zero as $\ell(\ell+1)/r^2+{\cal O}(1/r^3)$.
Hence, it necessarily has at least one maximum
and one inflection point.

On the other hand, the second term,
\begin{equation}
 F(r):=-G(r) \frac{\Box r}{r},
\end{equation}
is independent of the mode number $\ell$ and it
encodes the curvature of the manifold ${\cal M}_2$.
In fact, it is straightforward to write this term
as a simple linear combination of the curvature scalars \eqref{ricci} and \eqref{psi2},
since
\begin{equation}
\frac{\Box r}{r}={-}2\left(\Psi_2+\frac{^{(4)}R}{12}\right).
\end{equation}
In order to intuitively understand the curvature term $F(r)$, it is interesting
to see that, making use of \eqref{eq.boxtortoise}, it can be given in terms of the gradient of the function $n(r)$ used in Eq.~\eqref{eq.tortoise}
to define the tortoise coordinate, that is,
\begin{equation}
 F(r)=\frac{1}{r^2}\frac{\mbox{d}^2 r}{ \mbox{d} r_*^2}=\frac{1}{2r}(n^2(r))'.
\end{equation}
Note that, from \eqref{flatmetric}, the tortoise coordinate $r_*$ is defined such that 
radial null rays move at unit coordinate speed, that is, $(\mbox{d}r_*/\mbox{d}t)^2=1$.
Since, in terms of the areal radius function, null rays have a coordinate
speed given by $n^2(r):=(\mbox{d}r/\mbox{d}t)^2=(\mbox{d}r/\mbox{d}r_*)^2(\mbox{d}r_*/\mbox{d}t)^2=(\mbox{d}r/\mbox{d}r_*)^2$,
the term $n^2(r)$ can be understood as an effective refractive index.
Its gradient thus encodes the inhomogeneity of the medium,
and produces reflection and scattering of the waves.

Following the parameterization of the metric \eqref{eq:metrikh123}, it is easy to see that, in terms of the
shape functions, the effective refractive index reads $n(r)=-G(r)\sqrt{h_3(r)}$, and the Regge-Wheeler potential
takes the explicit form,
\begin{equation}\label{potentialGh3}
V_\ell(r)
=-G(r)\frac{l(l+1)}{r^2}+\frac{1}{2 r}\left(G^2(r)h_3(r) \right)'.
\end{equation}
Therefore, for the family of geometries under consideration,
$n(r)$ vanishes at the horizon,
it is positive definite for all the domain $r\in(r_H,\infty)$, and, with the assumed fall-off conditions
\eqref{eq.decays}, it tends to $n(r)=1+{\cal O}(1/r)$ at $r\to\infty$.
These properties imply that the curvature term $F(r)$ is zero at the horizon, while, for $r\to\infty$, $F(r)={\cal O}(1/r^3)$.
If we additionally assume that $n(r)$ increases monotonically with $r$,\footnote{If one has in mind a black-hole spacetime, this is a very mild
assumption, since it just implies that, as one goes further away
from the horizon, the gravitational pull decreases monotonically.
More specifically, this assumption implies that
the tortoise coordinate monotonically approaches
the areal radius function $r$, i.e., ${\rm d}r/{\rm d}r_*\to 1$,
and thus null geodesics monotonically tend to the ones
corresponding to a flat geometry.}
then $F(r)$ is positive definite for all $r\in(r_H,\infty)$,
and thus it has at least one maximum and one inflection point.

Therefore, both terms in the potential, $-G(r)\frac{\ell(\ell+1)}{r^2}$
and $F(r)$, feature the same qualitative form,
being zero both at the horizon and at asymptotic infinity,
and, under the assumption of monotonicity of $n(r)$, being both
positive definite for all the domain $r\in(r_H,\infty)$.
In the standard cases, such as the slightly deformed Schwarzschild
geometries we will consider in Secs.~\ref{sec_deformedschwarzschild} and \ref{sec:apltpbhsts}, each term has exactly one maximum,
and presents the characteristic hilltop shape.
The maximum and inflection point of each term do not coincide in general,
but their sum produces the typical potential barrier with
one maximum and a decaying behavior towards both ends
$r\to r_H$ and $r\to\infty$ (see Fig.~\ref{fig:potentials}).
For example, for the particular case of the Schwarzschild geometry,
$n(r)=-G(r)=(1-2M/r)$,
so the potential takes the explicit form\footnote{Here, and in the
following, we will use $\overset{\circ}{\mbox{}}$
to indicate quantities corresponding to
the Schwarzschild geometry with the value of the shape functions $h_1=1$,
$h_2=r$, and $h_3=1$, that is, for any $X$ we define,
$\overset{\circ}X:=X{\big|}_{h_1=1,h_2=r,h_3=1}$.}
\begin{align}
    \overset{\circ}{V}_{\ell}(r):= \left(1-\frac{2M}{r}\right)\left(\frac{\ell(\ell+1)}{r^2}+\frac{2M}{r^3}\right),
\end{align}
with the commented asymptotic behavior and one maximum located at $r=\overset{\circ}{r}_{\rm max}:=2M
(\frac43+f(\ell))$, see Eq.~\eqref{eq.xmax0} below.
However, geometries with stronger deviations from the Schwarzschild
geometry than the ones we will consider may present
several critical points.

\section{Radiative properties of the horizon}\label{sec.radiation}

As it is well known, black-hole horizons radiate as black bodies.
However, as shown above, in such backgrounds the field modes
do not propagate freely; they instead interact with the potential barrier
and get reflected and scattered.
Thus the radiation spectrum detected by a distant
observer differs from the actual spectrum radiated by the horizon.
More precisely, the radiation detection rate observed at large distances
takes the form
\begin{equation}\label{eq.gammad}
\Gamma_{\rm D}(\omega)= \frac{\sigma(\omega)}{e^{\hbar \omega/T}-1}.
\end{equation}
This expression is composed by two terms. On the one hand,
$(e^{\hbar \omega/T}-1)^{-1}$ provides the thermal emission rate from a hot black body,
with $\hbar\omega$ being the energy of the mode and $T$ the temperature of the body. On the other hand,
the nonthermal term $\sigma(\omega)\in[0,1]$ is called the graybody factor.
This factor is $\sigma(\omega)=1$ for an ideal black body, while, in general,
$\sigma(\omega)<1$, which modulates the profile of the radiation and measures the
departure from a pure black-body spectrum. In the case of ideal black holes,
as the ones studied in this paper, the origin of a nontrivial $\sigma(\omega)$
is due to the presence of the potential \eqref{potential}
that produces scattering and reflection
of the modes emitted by the horizon. 

In this section we analyze the radiative spectrum
from a general spherical nondegenerate horizon as parameterized
by the line element \eqref{eq:metrikG-param}. In particular, in Sec.~\ref{sec.temperature}
we present the computation of the horizon temperature $T_H$, and in Sec.~\ref{sec.graybodygeneral}
we examine the graybody factor $\sigma(\omega)$.
Then, in Sec.~\ref{sec_deformedschwarzschild}, we will study the temperature and the graybody factor for slightly deformed Schwarzschild black holes, while in Sec.~\ref{sec:apltpbhsts} we will discuss these thermodynamic properties
for specific black-hole geometries.

\subsection{Temperature}\label{sec.temperature}

To derive the general form of the temperature, we use the Hamilton-Jacobi approach for the tunneling picture \cite{Parikh:1999mf,Shankaranarayanan:2000gb,DiCriscienzo:2007pcr,DiCriscienzo:2010vz,Vanzo:2011nd,Vanzo:2011wq,moretti2012state,Giavoni:2020gui}. In this technique, the field is given by the
Wentzel-Kramers-Brillouin (WKB) approximation,
\begin{equation}\label{eq:wkb-feld}
    \phi(x)=\phi_o\;e^{\frac{i}{\hbar}S_0},
\end{equation}
where $\phi_o\in\mathbb{R}$ is assumed to be a slowly varying amplitude,
and, for all practical purposes, $\phi_o$ is treated as a constant.
The tunneling amplitude compares the incident with the transmitted intensity of the field,
\begin{equation}
    \frac{|\phi_{\rm trans}|^2}{|\phi_{\rm inc}|^2}\propto e^{-\frac{2}{\hbar}{\rm Im}(S_0)}.
\end{equation}
The tunneling rate by itself obscures
quantum field-theoretic features,
which resurface once the distributional nature
of the emission spectrum is made explicit through a comparison with a Boltzmann factor $e^{-\hbar\omega/T_H}$. The Hawking effect can then be defined through Im$(S_0)>0$, presumed Im$(S_0)\propto \hbar\omega$ \cite{Giavoni:2020gui}.
This yields the Hawking temperature 
\begin{equation}\label{eq:HawkT}
    T_H=\frac{\hbar \omega}{2\,{\rm Im}(S_0)}.
\end{equation}

To determine $S_0$, we first insert \eqref{eq:wkb-feld} into the Klein-Gordon equation ${}^{(4)}\Box\phi=0$, expand in powers of $\hbar$, and obtain, at leading order, 
\begin{equation}\label{eq.hamiltonjacobi}
 {}^{(4)}g(\nabla S_0, \nabla S_0)=0.
\end{equation}
This is the Hamilton-Jacobi equation for a massless particle,
which defines null geodesics, and
thus $S_0$ is its corresponding classical action.
In order to find a solution,
we will use an ansatz of the form $S_0=\int$d$S_0$, as well as the Hamilton equations to define the momenta $k=\mathcal{L}_VS_0$ along a vector field $V$.
Determining these momenta $k$ is related with the choice of a particular frame.
Here, for instance, we will use the Killing frame,
such that we can identify a conserved energy through the Hamilton equations.

More specifically, assuming for simplicity
a spherically symmetric configuration $S_0=S_0(t,r)$,
for the diagonal line element \eqref{eq:metrikG-param}
equation \eqref{eq.hamiltonjacobi} simply reads
\begin{equation}\label{eq.sphericalhamiltonjacobi}
(\partial_t S_0)^2= n^2 (\partial_r S_0)^2,
 \end{equation}
with $n=-G\sqrt{h_3}$, as defined above.
Plugging the ansatz in the Killing frame, that is,
\begin{equation}
 S_0=-\omega t+ \int_{r_H}^\infty \mbox{d}r\, k(r),
\end{equation}
into the Hamilton-Jacobi equation \eqref{eq.sphericalhamiltonjacobi} leads to
\begin{equation}
    k(r)=\pm \frac{\omega}{n(r)},
\end{equation}
where $\omega$ is the conserved frequency associated to the Killing vector $\partial_t$,
and the global sign of $k$ defines ingoing ($+$) or outgoing ($-$) modes. For definiteness,
let us choose an outgoing mode for the following discussion.
Furthermore, since we have assumed a nondegenerate horizon,
with $G(r_H)=0$ and $G'(r_H)<0$, and $h_3(r)$ to be positive definite,
the function $n(r)=-G\sqrt{h_3}$ has a simple zero at the horizon, i.e.,
$n(r)\sim n'(r_H)(r-r_H)+\mathcal{O}((r-r_H)^2)$, with $n'(r_H)>0$.
Therefore, $k(r)$ presents a simple pole there\footnote{
While our assumptions only allow for a simple pole, which is associated with any kind of simple trapping horizon,
in general there may occur other types of poles, e.g., at degenerate horizons
like in extremal Reissner-Nordström black holes. Nevertheless, these poles create no imaginary part for $S_0$.
Thus, by definition, the temperature of such horizons is zero and they produce no Hawking effect.} \cite{Giavoni:2020gui,Sebastiani:2018ktb}.
Hence, we define the above integral using the causal prescription
$n(r)\to n(r)- i 0$, which shifts the zero of $n(r)$ to the complex plane.
Then, making use of the standard distributional equality,
\begin{equation}
 \frac{1}{n(r)\pm i 0}={\rm PV}\left(\frac{1}{n(r)} \right) \mp i \pi\delta(n(r)),
\end{equation}
where `PV' denotes the Cauchy principal value, the imaginary part of $S_0$ reads
\begin{equation}
 {\rm Im}(S_0)= {\rm Im}\left(\int \mbox{d}r\, \frac{-\omega}{n(r)- i 0} \right)
 =-\pi \omega \int \mbox{d}r\, \delta(n(r))=-\frac{\pi \omega}{n'(r_H)}.
\end{equation}
From here it is thus straightforward to read off the temperature,
\begin{equation}
 T_H=-\frac{\hbar}{2 \pi} n'(r_H),
\end{equation}
which, up to overall constant factors, is completely determined by the gradient of $n(r)$ evaluated at
the horizon. We note that, as defined, the temperature is positive
since we assumed $G'(r_H)<0$, and thus $n'(r_H)<0$. In terms of the shape functions $n(r)=-G(r)\sqrt{h_3(r)}$, and
given that $h_3$ is positive definite and does not vanish at the horizon,
the temperature reads
\begin{equation}\label{eq:temperatur}
 T_H=-\frac{\hbar}{2 \pi} G'(r_H) \sqrt{h_3(r_H)}.
\end{equation}
This shows an interesting distinction between the shape functions.
While $G(r)$, and thus $h_1(r)$ and $h_2(r)$, appears as a derivative in the temperature,
$h_3(r)$ just contributes as a global multiplicative factor.

\subsection{Graybody factor}\label{sec.graybodygeneral}

In this section, we study how the potential influences the reflection coefficient and, as such, how it shapes the emitted spectrum through the graybody factor. As explained above, the graybody factor $\sigma_\ell(\omega)$
describes the scattering of the modes from the gravitational potential in the exterior region of the black hole.
Since the metric is static, the timelike Killing field allows us to perform a Fourier transformation of the modes $\psi_{\ell}(t,r_*)\sim u_{\ell}(r_*;\omega) e^{-i\omega t}$. In this way, the wave equation \eqref{eq:RWG}
is transformed into an ordinary second-order differential equation,
\begin{equation}\label{eq.mode}
\frac{{\rm d}^2 u_{\ell}(r;\omega)}{{\rm d}r_*^2}+V_\ell(r) u_{\ell}(r;\omega) =\omega^2\, u_{\ell}(r;\omega),
\end{equation}
that is reminiscent of the stationary Schr\"odinger equation for a particle in a static potential $V_\ell(r)$.
For convenience, we use the radius of the horizon $r_H$ to construct a
dimensionless radial distance,
\begin{equation}
 x:=r/r_H.
\end{equation}
Similarly, we define the dimensionless counterparts for the frequency $\nu:= \omega r_H$,
as well as for the potential
\begin{equation}
v_\ell(x):=V_\ell(r(x)) r_H^2,
\end{equation}
which, considering the potential in terms of the shape functions \eqref{potentialGh3},
takes the form
\begin{equation}\label{defvl}
v_\ell(x)=-G(x)\frac{\ell(\ell+1)}{x^2}+\frac{1}{2x}\frac{\mbox{d}}{\mbox{d}x} \Big[G^2(x) h_3(x) \Big].
\end{equation}
In this way, the mode equation \eqref{eq.mode} reads
\begin{equation}\label{eq.dimensionlessRWeq}
\frac{{\rm d}^2 u_{\ell}(x;\nu)}{{\rm d}x_*^2}+v_\ell(x) u_{\ell}(x;\nu) =\nu^2\, u_{\ell}(x;\nu),
\end{equation}
with $x_*:=r_*/r_H$.
As explained in the previous section, generically the Regge-Wheeler potential $v_\ell$ is vanishing both at the horizon and at infinity, and has at least one maximum.
Standard cases, such as the slightly deformed black holes we will study below, have
one exact maximum and present the typical hilltop shape.
However, in principle, general shape functions could create multiple maxima in the potential
and each thereof would contribute to the graybody factor.
But, since we will effectively approximate the potential
by a parabola, the methodology we will use for the analytic
derivations can only be applied to one maximum.\footnote{If we were to face multiple maxima for a given frequency $\omega$,
we would get a product of individual graybody factors, i.e., $\sigma_\ell(\omega)\sim\prod_i\sigma_\ell^i(\omega)$,
as a rough approximation for the total graybody factor.}

Therefore, by assumption, the potential only has one global maximum, which is located at $x_{\rm max}:={v'_\ell}^{-1}(0)$.
For a mode with given frequency $\nu$, there are two main regimes related to the sign of
the WKB frequency $v_\ell(x)-\nu^2$.
On the one hand, in case the maximum of the potential $v_\ell(x_{ \rm max})$
is lower than the squared frequency, that is, $\nu^2>v_\ell(x_{ \rm max})$,
then the WKB frequency is negative,
with the result that the potential barrier becomes almost transparent for the mode.
As such, under the present approximation, the graybody factor for
$\nu^2>v_\ell(x_{\rm max})$ is $\sigma_\ell(\nu)\approx1$.

On the other hand, the nontrivial case is $\nu^2<v_\ell(x_{\rm max})$,
because for such mode the barrier turns opaque. In this case,
the WKB frequency admits two turning points $x=\xi$, that is, two
real zeros of the combination $(v_\ell(\xi)-\nu^2)=0$. 
Then, for frequencies below the maximum of the potential, that is, $\nu^2<v_\ell(x_{\rm max})$,
one can use the WKB approximation to write the solution to the
equation \eqref{eq.dimensionlessRWeq} as
\begin{equation}\label{eq:WKBFrequ}
u_\ell(x;\nu)=\frac{C_+\exp\left(\int_{\xi_-}^{\xi_+}\frac{{\rm d}x}{n(x)}\sqrt{v_\ell(x)-\nu^2}\right)+C_-\exp\left(-\int_{\xi_-}^{\xi_+}\frac{{\rm d}x}{n(x)}\sqrt{v_\ell(x)-\nu^2}\right)}{\sqrt[4]{v_\ell(x)-\nu^2}},
\end{equation}
with two integration constants $C_\pm$,
and the two turning points $\xi_{\pm}$ defined such that $\xi_{-}<x_{\rm max}<\xi_{+}$.
In order to define the graybody factor, we choose the positive-frequency mode, i.e.,
$C_-\equiv0$ and $C_+\equiv1$, such that it coincides with the positive-frequency mode in the limit of Minkowski spacetime.

Since the turning points $\xi_\pm$ are not always easy to determine analytically, Iyer and Will \cite{Iyer:1986np} developed a versatile method that approximates the Regge-Wheeler potential iteratively through polynomials around the potential's maximum. At first order, the approximation uses an inverted parabola, but higher-order terms can be calculated straightforwardly \cite{Iyer:1986np}.
Based on this, Konoplya and Zhidenko provided a method to determine the graybody factor directly \cite{Konoplya:2019hlu}.
By definition, $\sigma_\ell(\nu)$ equals the transmission coefficient for outgoing modes
$|t_{\nu\ell}|^2$, which, making use of the completeness relation between $|t_{\nu\ell}|^2$
and the reflection coefficient $|r_{\nu\ell}|^2$, can be written as
\begin{equation}\label{eq:gkf}
\sigma_\ell(\nu):=|t_{\nu\ell}|^2=1-|r_{\nu\ell}|^2=\frac{1}{1+e^{2\pi K_\ell(\nu)}}.
\end{equation}
The exponent function $K_\ell(\nu)$ is defined via the WKB frequency as
\begin{equation}\label{eq:KKon}
    K_\ell(\nu)=\frac{v_\ell(x_{\rm max})-\nu^2}{\sqrt{-2v_\ell''(x_{\rm max})}},
\end{equation}
where $v''_\ell(x_{\rm max})<0$ denotes the second derivative of $v_\ell(x)$ evaluated at $x_{\rm max}$.
Therefore, within this approximation,
the graybody factor for a given frequency $\nu$ is fully specified by the two quantities
$v_\ell(x_{\rm max})$ and $v''_\ell(x_{\rm max})$.
The value $v_\ell(x_{\rm max})$ corresponds to the height
of the potential at its maximum, while
$v''_\ell(x_{\rm max})$ encodes its curvature (i.e.,
the width of the barrier) around $x=x_{\rm max}$.
Consequently, as expected, a taller and broader potential barrier,
corresponding to a higher $v_\ell(x_{\rm max})$ and a lower $|v''_\ell(x_{\rm max})|$, suppresses the tunneling probability,
so that fewer modes are transmitted, and
the value of the graybody factor decreases.

Finally, it is important to mention that this approximation comes with a certain range of validity
determined by the distance,
\begin{equation}\label{eq:gkb-wkb}
  d^\ell_{\rm max}(\nu)= \max_{\xi_-\le x\le \xi_+}|x-x_{\rm max}|<\sqrt{\frac{2(\nu^2-v_\ell(x_{\rm max}))}{v_\ell''(x_{\rm max})} },
\end{equation}
from the turning points $\xi_{\pm}$ to the location of
the maximum of the potential $x_{\rm max}$ \cite{Konoplya:2019hlu}.
This distance depends on the values of $\ell$ and $\nu$, and,
effectively, $d^\ell_{\rm max}(\nu)$ determines the digression of the potential from a parabola.
The approximation is thus valid for those combinations of
$\ell$ and $\nu$ with a relatively small value of $d^\ell_{\rm max}(\nu)$.
Intuitively this implies frequencies
$\nu^2<v_\ell(x_{\rm max})$ that are close to $v_\ell(x_{\rm max})$.
Frequencies larger than the potential height $\nu^2>v_\ell(x_{\rm max})$ would
define an imaginary $d_{\rm max}^\ell$, since, as commented above, this computation does not apply to such
large frequencies and, within the present approximation, the graybody factor is considered to be
approximately 1 for them.

In summary, under the present approximation, one finds that:
\begin{itemize}
\item For modes with high frequencies, that is, $\nu^2>v_\ell(x_{\rm max})$,
the potential barrier is transparent and thus
the graybody factor is $\sigma_\ell(\nu)\approx 1$.
\item Modes with frequencies
much lower than the maximum $\nu^2\ll v_\ell(x_{\rm max})$,
are completely blocked by the potential barrier, and thus
the graybody factor is $\sigma_\ell(\nu)\approx 0$.
\item For modes with intermediate frequencies, that is
$\nu^2< v_\ell(x_{\rm max})$, but close to the maximum $\nu^2\sim v_\ell(x_{\rm max})$,
the graybody factor is given by \eqref{eq:gkf}
with the exponent function \eqref{eq:KKon}.
\end{itemize}
The latter one is the nontrivial case that we will study
in detail in the following sections for slightly modified
Schwarzschild geometries.

\section{Radiation of slightly deformed Schwarzschild geometries}\label{sec_deformedschwarzschild}

After analyzing the radiative properties of the outermost horizon of a generic spherical black hole,
in this section we will specifically consider small deformations of the Schwarzschild geometry
in order to obtain its corresponding temperature and graybody factor.
We begin by considering the parameterization \eqref{eq:metrikG-param} for a static spherical geometry,
in terms of the functions $G(r)$ and $h_3(r)$, and we recall that the Schwarzschild geometry corresponds
to the specific forms $G(r)=-(1-2M/r)$ and $h_3(r)=1$. Therefore, in order to construct
a slightly deformed Schwarzschild geometry, we first define
the squared norm of the Killing as
\begin{equation}\label{defG}
 G(r)=-\left(1-\frac{2M}{r}\right)- \epsilon \left(\frac{2 M}{r}\right)^j,
\end{equation}
with a dimensionless parameter $|\epsilon|\ll 1$ and a power $j\geq 2$, which ensures
that asymptotically $G$ tends to its Schwarzschild form.
Therefore, we are considering a modification to Schwarzschild as a power $1/r^{j}$,
which is strongest at the horizon and becomes negligible as we tend to large
values of $r$.

In order to perform a meaningful comparison between the deformed
geometry and its Schwarzschild counterpart, it is necessary to decide which property
is kept constant. At first sight, it seems natural to keep the parameter $M$
constant and compare ``black holes of equal mass''. However, as explained above,
even if the parameter $M$ is related to the mass of the black hole,
depending the definition of mass, it acquires certain modifications.
Therefore, instead of considering black holes with
the same mass, we will be comparing ``black holes of the same size'', i.e.,
with the same horizon area $4\pi r_H^2$. This is a well defined
geometric quantity, that in principle could also be observationally measured,
and it is completely determined by $r_H$, which is
the largest root of $G(r_H)=0$.
Therefore, in the following, we will be using the position of the horizon
$r_H$ as a constant and fixed parameter.

Going back to the definition \eqref{defG}, it is easy to see that
the position of the horizon is given by
\begin{equation}
 r_H=2 M\left(1-\epsilon+(1-j) \epsilon^2\right)+{\cal O}(\epsilon^3).
\end{equation}
From now on, since we are assuming $|\epsilon|\ll 1$, in general
we will linearize all expressions in $\epsilon$, and drop all higher-order powers.
In this way, we can write $2M=r_H(1+\epsilon)$, and thus,
in terms of the horizon radius $r_H$, the definition \eqref{defG} takes the form,
\begin{equation}\label{defGintermsofrh}
 G(r)=-1+(1+\epsilon)\frac{r_H}{r}- \epsilon \left(\frac{r_H}{r}\right)^j.
\end{equation}
In addition, in order to completely characterize the metric, we introduce
another two dimensionless parameters $\lambda$ and $k$ to encode the modifications
of the remaining shape function,
\begin{equation}\label{defh3}
h_3(r)=1-\lambda\left(\frac{r_H}{r} \right)^k ,
\end{equation}
with $|\lambda|\ll1$ and $k\geq 1$. Similarly as with $\epsilon$, we will also
consider quadratic and higher-order powers of $\lambda$ to be negligible.

Note that, in terms of the dimensionless radial coordinate defined above $x=r/r_H$, the metric functions simply read,
\begin{align}\label{gintermsofepsilon}
 G(x) &=-1+\frac{(1+\epsilon)}{x}- \frac{\epsilon}{x^j},\\\label{h3intermsoflambda}
h_3(x) &=1 -\frac{\lambda}{x^k},
\end{align}
and thus the explicit dependence on $r_H$ is absorbed in $x$.
Even if this ansatz of the shape functions may seem very restrictive,
as will be made explicit in Sec.~\ref{sec:apltpbhsts},
many well-known black-hole models in the literature
fit in this description.
Some models present exactly this form of the metric functions, with power-law corrections to the Schwarzschild
form, while others must be Taylor expanded in
their corresponding smallness parameters $\epsilon$ and $\lambda$
to take this form. Thus, typically
$j$ and $k$ will be integer numbers.
This is not a requirement for the validity of
our results. However, in order to provide some examples below, we will indeed assume integer values of $j$ and $k$.

As a final note, let us recall that we are assuming the existence
of an outermost nondegenerate horizon with a static exterior.
As defined in \eqref{gintermsofepsilon},
$x=1$ is, by construction, a root of $G(x)$, but $G'(1)<0$
only if $\epsilon<1/(j-1)$. This condition also ensures there
are no other roots of $G(x)$ for $x>1$, and that $G(x)$ is asymptotically
decreasing in $x\in(1,\infty)$. In addition, the requirement
of a Lorentzian spacetime implies $h_3(x)>0$, which is obeyed
in the range $x\in(1,\infty)$ only if $\lambda\leq 1$.
Therefore, the
spacetime structure we are assuming imposes the bounds
$\epsilon<1/(j-1)$ and $\lambda\leq 1$ in the deformation
parameters. In principle, these are not strong bounds, since we are indeed
assuming both parameters to be small in magnitude,
that is, $|\epsilon|\ll 1$ and $|\lambda|\ll 1$, though
the condition $\epsilon<1/(j-1)$ may be relevant if
one considers very large values of $j$.

\subsection{Temperature}

For such a deformation of the Schwarzschild geometry, we can easily read off the temperature from \eqref{eq:temperatur} to be
\begin{equation}
    T_H=\frac{\hbar}{4\pi r_H}\sqrt{1-\lambda}(1-(j-1)\epsilon).
\end{equation}
This expression is exact, but if we
linearize it in the smallness parameters $\epsilon$ and $\lambda$,
it reads, $$T_H=\overset{\circ}T_{H}+\delta  T_H,$$ with 
\begin{equation}\label{eq.schtemperature}
\overset{\circ}T_{H}:=\frac{\hbar}{4\pi r_H}
\end{equation}
being the temperature of the Schwarzschild horizon and
\begin{equation}\label{eq.linearizedTcorrection}
    \delta T_H:= -\frac{\hbar}{8\pi r_H}(\lambda+2(j-1)\epsilon).
\end{equation}
Therefore, it is straightforward to see that negative $\lambda$ and $\epsilon$ increase the temperature, while positive $\lambda$ and $\epsilon$ decrease it
(recall that we are assuming $j\geq 2$). It is noteworthy to say that the particular power
$k$ of the $\lambda$-correction does not enter the temperature, because, as commented above
[see Eq.~\eqref{eq:temperatur}], $h_3(x)$ appears only as a global prefactor
evaluated at the horizon $x=1$, and not as a derivative. 

\subsection{Graybody factor}

As explicitly shown in Eqs.~\eqref{eq:gkf}--\eqref{eq:KKon}, the main contribution to the graybody factor is given by the height and width of the potential barrier.
Therefore, before providing analytic expressions for
the graybody factor of the deformed Schwarzschild geometry in Sec.~\ref{sec.graybody},
we will first, in Sec.~\ref{sec.shape}, analyze in detail
how the shape of the Regge-Wheeler potential depends on the parameters $\lambda$ and $\epsilon$.
 However, Sec.~\ref{sec.shape} can be skipped without losing the main line of discussion,
and the interested reader can proceed directly to Sec.~\ref{sec.graybody}.

\subsubsection{The shape of the potential barrier}\label{sec.shape}

\begin{figure}
       \centering
           \includegraphics[width=\linewidth]{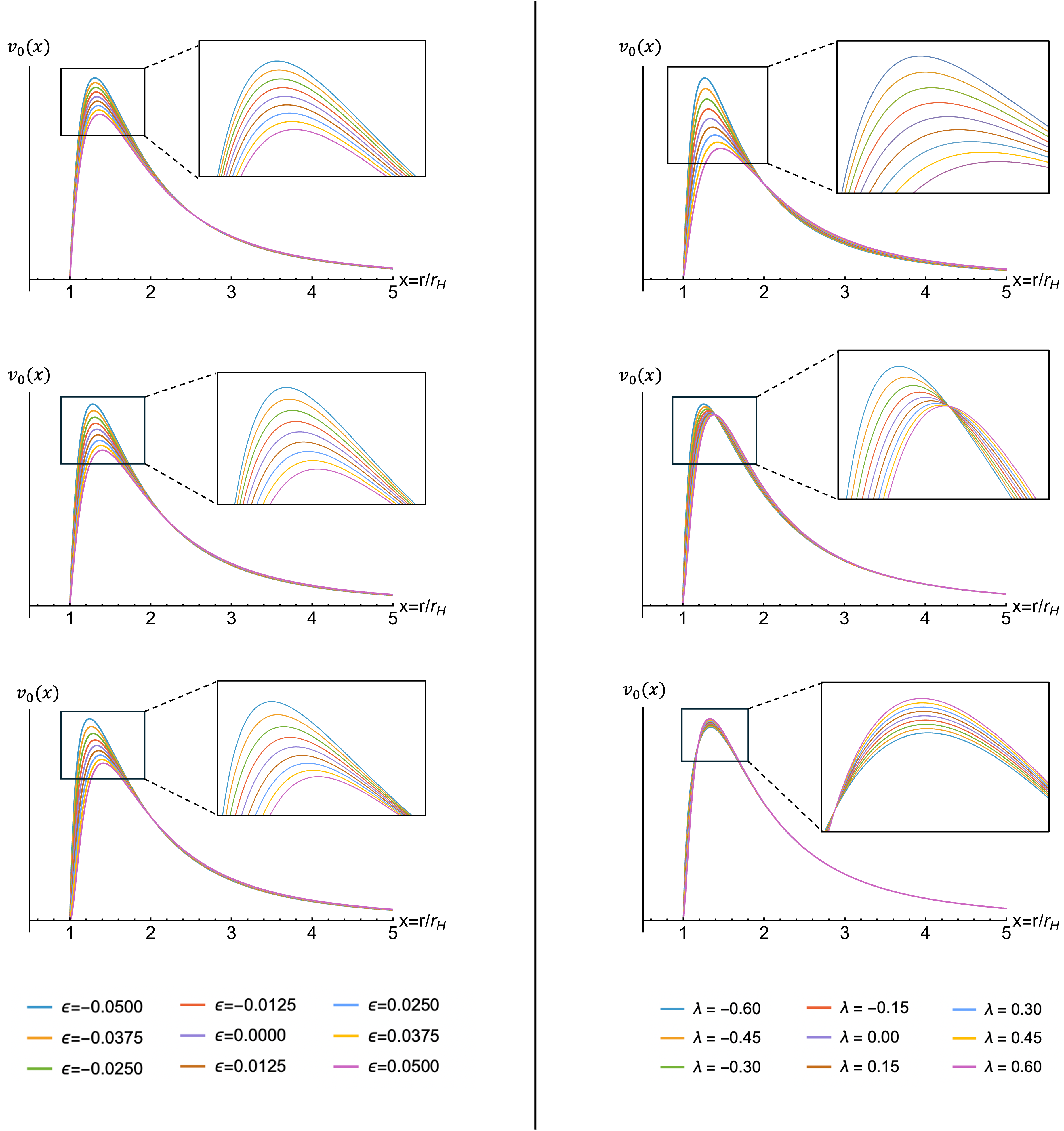}
            \begin{minipage}{0.45\linewidth}
        \small
        \textbf{(a)} Dimensionless potential $v_0(x)$ for $\lambda=0$ and different values of $\epsilon$. The power of the corrections is
        fixed for each plot, with $j=3,5$, and $11$ in the top, middle,
        and bottom panels, respectively. In general, larger values of $\epsilon$ produce smaller potential barriers. In the zoomed-in regions, one can appreciate how the maximum shifts to the right (further away from the horizon) as $\epsilon$ increases.
    \end{minipage}
    \hfill
    \begin{minipage}{0.45\linewidth}
        \small
        \textbf{(b)} Dimensionless potential $v_0(x)$ for $\epsilon=0$ and different values of $\lambda$. The power of the corrections is fixed
        in each plot, with $k=2,5$, and 11 in the top, middle, and bottom panels, respectively. Note that the change on the height of the potential with respect to $\lambda$ is different for small and large powers $k$. In the former case (top and middle panels), a larger $\lambda$ implies a  smaller potential, whereas in the latter case (bottom panel), a larger $\lambda$ implies a larger potential.
    \end{minipage}
        \caption{Shape of the potential barrier for s-waves ($\ell=0$) for different values of the correction parameters $\epsilon$, $\lambda$,
        $j$, and $k$.}
        \label{fig:potentials}
\end{figure}

In contrast to the temperature, which is defined locally at the horizon,
the Regge-Wheeler potential spans all the exterior region $r>r_H$.
Replacing expressions \eqref{gintermsofepsilon}--\eqref{h3intermsoflambda} in the dimensionless potential \eqref{defvl}, and linearizing
in both $\epsilon$ and $\lambda$, we obtain
\begin{align}\label{eq.potlin}
v_\ell(x)=\overset{\circ}v_\ell(x)
-\frac{\lambda}{x^{4+k}}(x-1)\left(1-\frac{k}{2}(x-1)\right)
+\frac{\epsilon}{x^3}\left(1-\ell(\ell+1)-\frac{2}{x}+\frac{\ell(\ell+1)-j}{x^{j-1}}
+\frac{j+1}{x^{j}}\right),
\end{align}
with
\begin{equation}\label{eq:potS}
\overset{\circ}v_\ell(x):=\frac{x-1}{x^4}(1+\ell(\ell+1)x),
\end{equation}
being the dimensionless potential for the Schwarzschild black hole.
As already commented above, for Schwarz\-schild, the potential is positive definite,
it vanishes both at the horizon $(x=1)$ and at the asymptotic infinity
$(x\to\infty)$, and it presents the characteristic
hilltop shape with exactly one maximum and one inflection point.
For relatively small values of the modification parameters
$|\epsilon|$ and $|\lambda|$, the qualitative form
of the potential is the same as for Schwarzschild. In particular,
for illustration, the potential \eqref{eq.potlin}
is plotted in Fig.~\ref{fig:potentials} for s-waves $(\ell=0)$,
which is the mode that typically carries the largest amount
of energy, and for certain choices of the parameters
$\epsilon$, $\lambda$, $j$, and $k$.
Before commenting these plots in more detail,
let us analytically show some of the modifications
produced by these parameters.

First of all, from expression \eqref{eq.potlin} it is clear that
the potential vanishes at the horizon ($x=1$), independently of
the value of the parameters $\epsilon$ and $\lambda$, as well as
of the mode number $\ell$,
and of the powers $j$ and $k$. Actually,
in the near-horizon regime, it is easy to obtain
\begin{equation}
v_\ell=(x-1)\left(1+\ell(\ell+1)-\lambda-\epsilon\, (j-1)(2+\ell(\ell+1))\right) +{\cal O}((x-1)^2).
\end{equation}
Therefore, a negative $\lambda$ and $\epsilon$
produce a steeper slope of the potential near the horizon, while
a positive $\lambda$ and $\epsilon$ define a flatter potential
(recall that $j\geq 2$ and $\ell\geq 0$).

Concerning the asymptotic end, as $x\to\infty$ the potential can be approximated by 
\begin{equation}
v_\ell=\frac{\ell(\ell+1)}{x^2} -\frac{(1+\epsilon)}{x^3}(\ell(\ell+1)-1) +{\cal O}\left(\frac{1}{x^4}\right).
\end{equation}
Thus, the dominant correction to the potential
far away from the horizon is given by the parameter $\epsilon$,
and it decays as $1/x^3$. On the one hand, for $\ell\neq 0$,
the decay rate of the correction is faster
than the decay rate $\sim 1/x^2$ of the dominant centrifugal term.
And the correction increases (for $\epsilon<0$) or decreases
(for $\epsilon>0$) the value of the potential. On the other hand,
for $\ell=0$, both the correction and the dominant term decay
with the same $\sim 1/x^3$ rate. For $\ell=0$, $\epsilon$
has the opposite effect as for the rest of the modes:
the correction produces a positive or negative contribution
for positive or negative $\epsilon$, respectively.
Concerning the correction parameterized by $\lambda$, its effects
decay very fast $\sim 1/x^{4+k}$ and have little impact
asymptotically.
As will be commented below in more detail, this is
quite a general feature, not only in the asymptotic regime. That is,
given the same order of magnitude of the parameters $\lambda$
and $\epsilon $, the modifications produced
by $\epsilon$ are typically stronger than those produced by $\lambda$.

\begin{figure}
        \centering
        \begin{subfigure}{\linewidth}
        \centering
        \includegraphics[width=.7\linewidth]{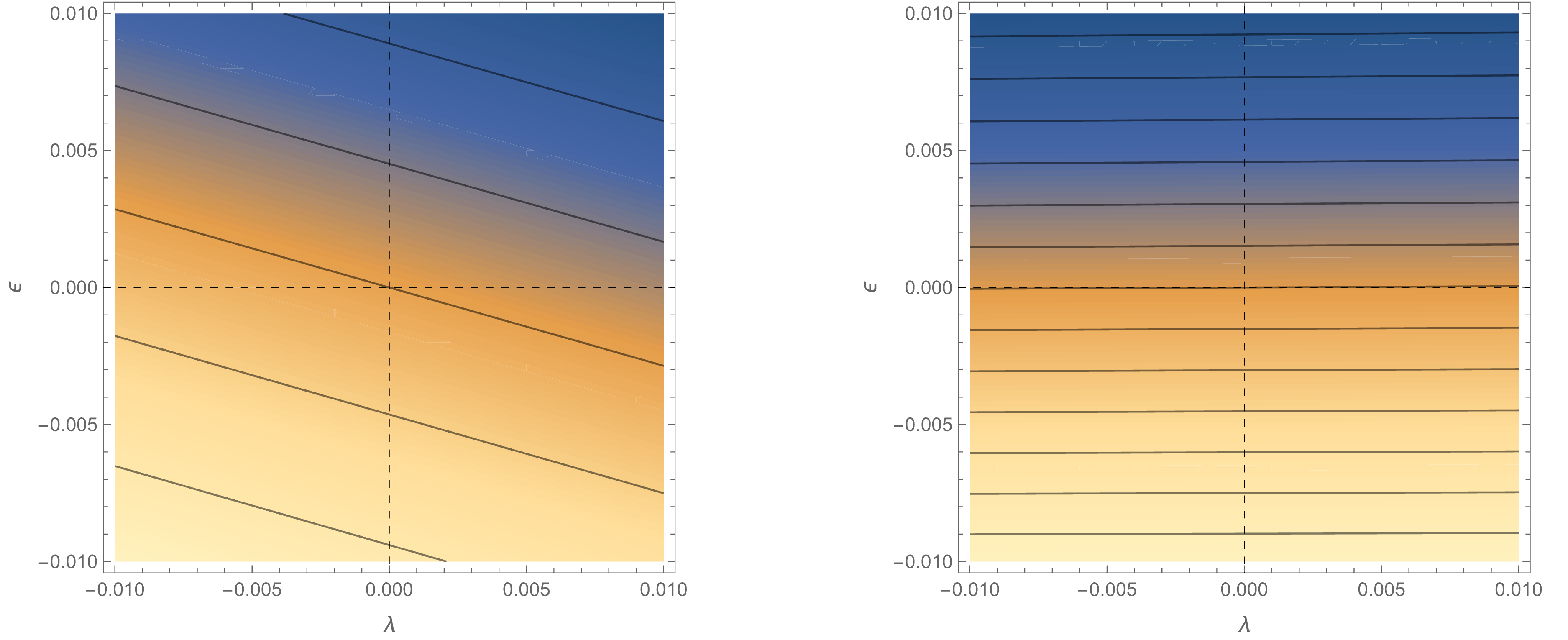}
        \caption{Position of the maximum of the potential $x_{\rm max}$. The parallel solid lines are contour lines at $\pm0.2\%$, $\pm0.4\%$, $\pm0.6\%$, $\pm0.8\%$, $\pm1.0\%$, and $\pm1.2\%$ deviations from the
        value corresponding to Schwarzschild.
        Both for small (left panel) and large (right panel) powers,
        $x_{\rm max}$ increases with increasing $\epsilon$.
        The effect of $\lambda$ changes from one case to the other,
        though for large powers its effects are much smaller than those produced
        by $\epsilon$.}
        \label{fig:contour_rmax}
        \end{subfigure}
        
        \vspace{12pt}
\begin{subfigure}{\linewidth}
        \centering
        \includegraphics[width=.7\linewidth]{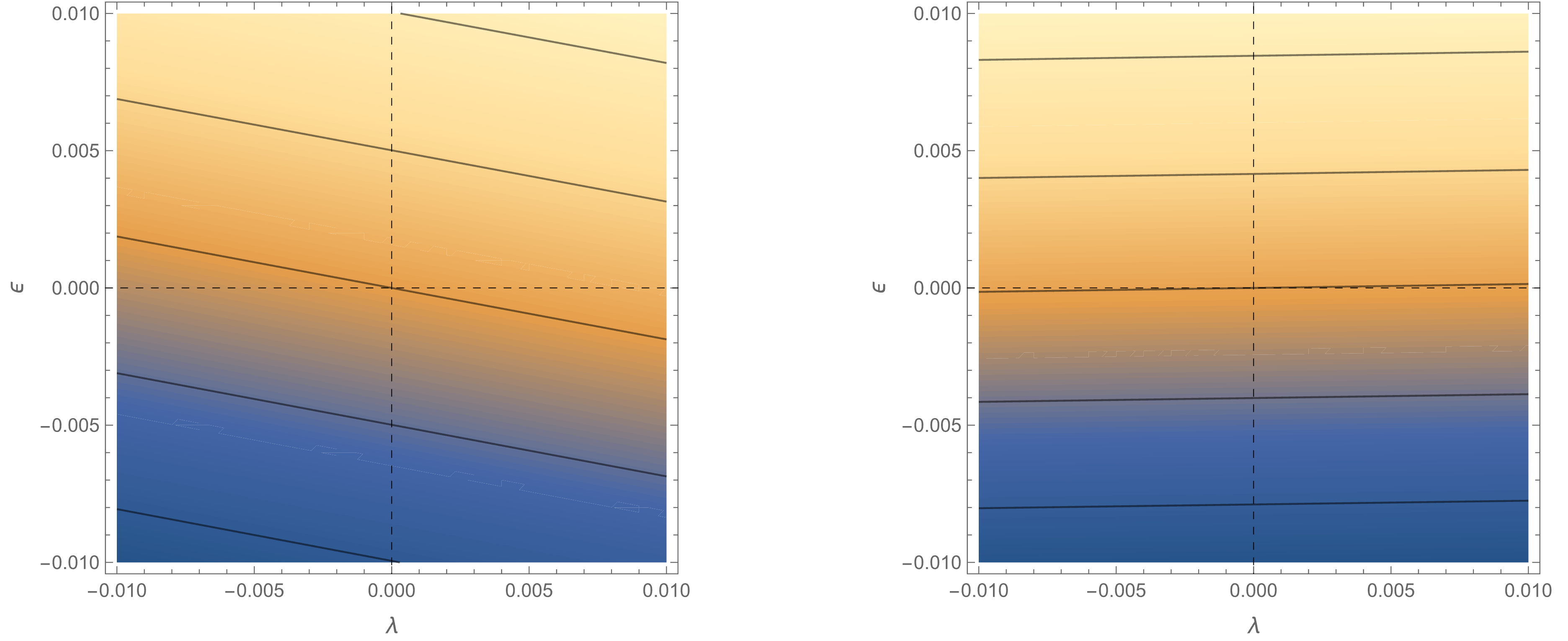}
        \caption{Height of the dimensionless potential barrier $v_0(x_{\rm max})$. The parallel black lines are contour lines at $\pm1\%$ and $\pm2\%$ deviations from the Schwarzschild value.
        The effects of $\lambda$ and $\epsilon$ are correlated
        for small values of $(j,k)$---plot on the left---, increasing
        $v_0(x_{\rm max})$ as $\epsilon$ and $\lambda$ decrease.
        The trend for $\lambda$ changes for large values of $k$,
        as can be seen comparing the left and right plots.}
        \label{fig:contour_vmax}
\end{subfigure}

\vspace{12pt}
\begin{subfigure}{\linewidth}
        \centering
        \includegraphics[width=.7\linewidth]{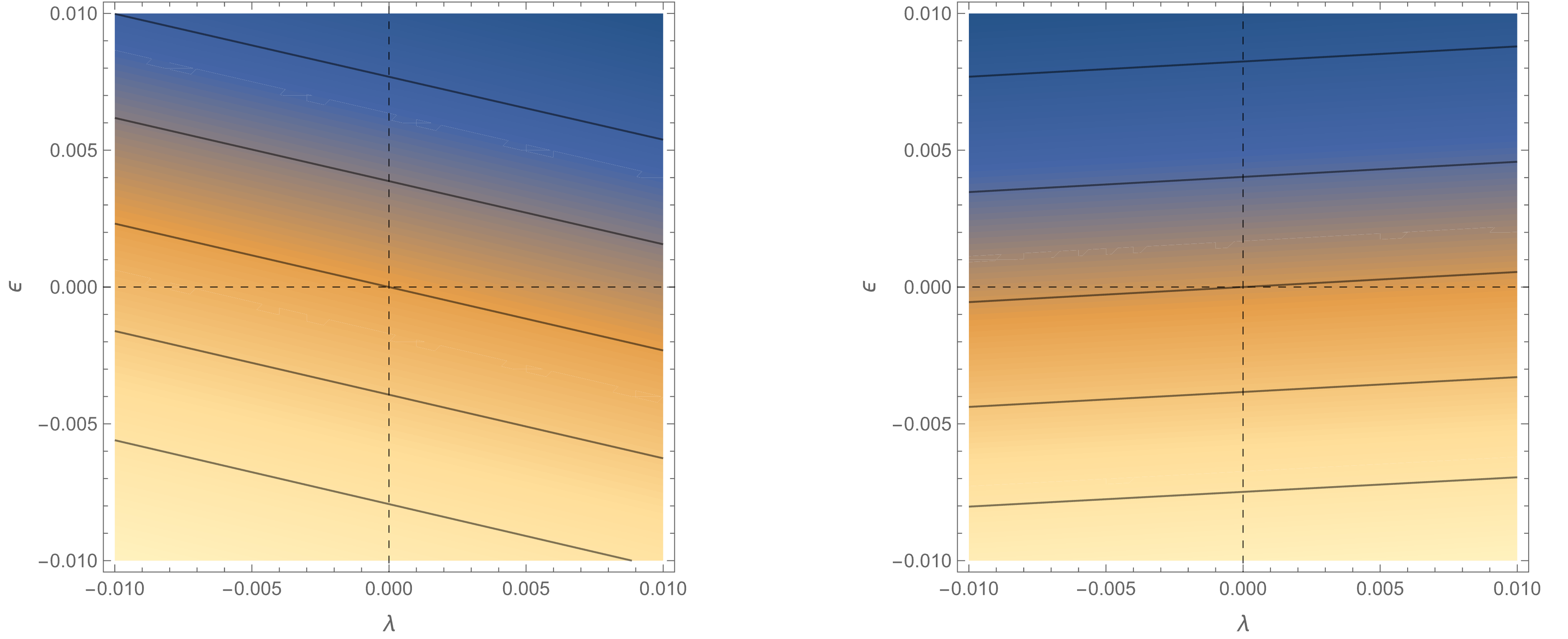}
        \caption{Width of the potential barrier $w=1/\sqrt{|v_0''(x_{\rm max})|}$. The parallel black lines are contour lines at $\pm1\%$ and $\pm2\%$ deviations from the Schwarzschild value.
        The behavior of $w$ in terms of $\epsilon$ and $\lambda$ is qualitatively
        similar as the behavior of $x_{\rm max}$ shown above.}
        \label{fig:contour_width}
\end{subfigure}
\caption{contour plots showing the behavior of different magnitudes of
the dimensionless potential $v_\ell(x)$
as functions of the parameters $\epsilon$ and $\lambda$.
All the plots correspond to s-waves ($\ell=0$)
and fixed values of the powers $j$ and $k$:
$(j,k)=(3,2)$ (for plots on the left column)
and $(j,k)=(11,11)$ (for the right column). These are
representative values to show the behavior of the different
magnitudes with small and large values of the powers $(j,k)$, respectively.
Lighter colors on the contour plots
correspond to smaller values of the plotted magnitude.
Dashed lines denote $\epsilon=0$ and $\lambda=0$.
Solid lines are contour lines with constant value of the magnitude under
consideration. In particular,
the solid line crossing the origin $(\epsilon=0=\lambda)$
corresponds to the value of the plotted magnitude
for the Schwarzschild geometry.}
\label{fig.contours}
\end{figure}

Regarding the global maximum of the potential, where $v'_\ell(x_{\rm max})=0$,
its position is given by
\begin{equation}\label{xmax}
 x_{\rm max}=\overset{\circ}x_{\rm max}+\delta x_{\rm max}.
\end{equation}
Here we have defined $\overset{\circ}x_{\rm max}$ to be the position of the maximum for the Schwarzschild geometry, which can be written as
\begin{equation}\label{eq.xmax0}
\overset{\circ}x_{\rm max}=\frac{4}{3}+f(\ell),
\end{equation}
with the function $f(\ell)$ given by
$$f(\ell)=-\frac{1}{12 \ell(\ell +1)}(9+7 \ell(\ell+1)-3\sqrt{9+\ell(\ell+1)(14+9 \ell(\ell+1))}).$$
This function is a monotonically increasing function of $\ell$,
with $f(0)=0$ and $\lim_{\ell\to\infty} f(\ell)=\frac{1}{6}$. Therefore,
the mode number $\ell$ pushes the position of the maximum of the potential
further away from the horizon, though it is always located in the interval $\overset{\circ}x_{\rm max}\in[\frac{4}{3},\frac{3}{2}]$.
The remaining term in \eqref{xmax}, $\delta x_{\rm max}$, is the displacement
of the position of the maximum produced by the corrections, and it reads,
\begin{equation}\label{eq:MaxVersch}
    \delta x_{\rm max}=\frac{ 2^{-2 k-1} 3^{k-2} (10-k) k}{4 \ell (\ell+1)+3}\lambda
    +\frac{4  \left(\left(\frac{3}{4}\right)^j \left(-4 \ell(\ell+1) (j+2)+(j-4) j-9\right)+9 \ell (\ell+1)+9\right)}{9 (4 \ell (\ell+1)+3)}\,\epsilon.
\end{equation}
As we can see here, the sign of $\delta x_{\rm max}$, and thus the direction of the displacement, highly depends on $j$, $k$, and $\ell$. Hence, a prediction of whether the corrections contribute 
to push the position of the maximum closer or further away
from the horizon nontrivially depends on these values. 
In order to showcase these effects,
Fig.~\ref{fig:contour_rmax} presents contour plots of
$x_{\rm max}$ as a function of $\epsilon$ and $\lambda$ for
s-waves $(\ell=0)$, and for two different couple of
values of $j$ and $k$.

Now that we have the position of the maximum at hand,
we can evaluate the height of the potential peak:
$$v_\ell(x_{\rm max})=\overset{\circ}{v}_\ell(x_{\rm max})+\epsilon\, \delta v^\epsilon_\ell(x_{\rm max})+\lambda\, \delta v^\lambda_\ell(x_{\rm max}),$$
where we have defined the height of the potential barrier for Schwarzschild,
\begin{align}
    \overset{\circ}{v}_\ell(x_{\rm max})&=\frac{12 \ell (\ell+1) (13 \ell (\ell+1)+18)+81}{256 (4 \ell (\ell+1)+3)},
\end{align}
and the correction terms coming from the presence of $\epsilon$ and $\lambda$ as functions of the parameters $j$ and $k$:
\begin{align}
    \delta v^\epsilon_\ell(x_{\rm max})&=\frac{6 \ell (\ell+1)\! \left(3^j\! \left(-4 \!\left(\ell(\ell+1)+4\right) j+40 \ell(\ell+1)+j^2+63\right)\!-9 (3 \ell (\ell+1)\!+4) 4^j\right)\!-2\ 3^{j+3} (j-3)-4^j 81}{2^{2 j+7} (4
   \ell (\ell+1)+3)},\\
   \delta v^\lambda_\ell(x_{\rm max})&=-\frac{ 3^{k+1} (\ell (\ell+1) (k-18) (k-4)-9 (k-6))}{2^{2 k+9}(4 \ell (\ell+1)+3)}.
\end{align}
The effects of these corrections can be seen in Fig.~\ref{fig:contour_vmax},
where we have plotted $v_\ell(x_{\rm max})$ as a function
of $\epsilon$ and $\lambda$
for $\ell=0$ and some specific values of $j$ and $k$.

Finally, the second derivative of the potential evaluated at the maximum
$v_\ell''(x_{\rm max})$ encodes the curvature of the potential barrier near the peak, and
the square root of its inverse $w:=1/\sqrt{|v_\ell''(x_{\rm max})|}$
can be interpreted as the width of the potential barrier.
In Fig.~\ref{fig:contour_width} we present contour plots
of $w$ as a function of $\epsilon$ and $\lambda$ for $\ell=0$
and fixed values of $j$ and $k$.

All the above expressions are quite involved, and the casuistry of the corrections is extensive, since the effects on the potential
vary significantly for each
specific set of values of the parameters $\ell$, $\epsilon$, $\lambda$, $j$,
and $k$. Let us, however, give some general trends,
which are visualized in the specific examples plotted in Figs.~\ref{fig:potentials} and \ref{fig.contours}.

Let us first focus on Fig.~\ref{fig:potentials}, where we plot the
dimensionless potential $v_{\ell}(x)$ as a function of $x$,
for s-waves ($\ell=0$) and fixed values of $\epsilon$, $\lambda$,
$j$, and $k$. More precisely, on the left column,
$\lambda$ is fixed to zero, and thus we exclusively observe the effects of $\epsilon$ on the potential barrier for different values of $j$.
In every case, an increase in $\epsilon$ flattens the peak, and moves it further away from the horizon. This effect is enhanced by the power of the correction $j$. Note that the inflection point of the curve is always around two horizon-radius ($x\approx 2$),
while the difference between curves corresponding
to different values of $\epsilon$ is suppressed
for larger values of $x$.

In contrast, the power $k$ of the $\lambda$-corrections changes the character of the modifications (see now the right column in Fig.~\ref{fig:potentials},
where $\epsilon=0$ and different values of $\lambda$ are shown).
Small values of $k$ (top and middle plots) show a flatter and smaller peak as $\lambda$ increases, with the difference being noticeable for small variations of $\lambda$. On the contrary, large values of $k$ (plot in the bottom) invert the effects of $\lambda$: for sufficiently big $k$, the peak of the potential becomes larger and sharper as we increase $\lambda$, showing the opposite behavior as in the small-$k$ scenario. This can be better appreciated in the zoomed-in regions of the
middle and bottom plots on the right column of Fig.~\ref{fig:potentials}, where one can see that the crossing point between curves moves to the left (closer to the horizon) as we increase $k$, and eventually goes to the left of the peak itself. 

In Fig.~\ref{fig.contours} we present contour plots showing
different magnitudes of the potential as a function of $\epsilon$
and $\lambda$, for $\ell=0$ and certain values of $j$ and $k$. More specifically, we plot
the location of the maximum $x_{\rm max}$ in Fig.~\ref{fig:contour_rmax},
the height of the peak of the potential $v_0(x_{\rm max})$ in Fig.~\ref{fig:contour_vmax},
and its width $w$ in Fig.~\ref{fig:contour_width}.
As one can appreciate in all these contour plots, in general,
given $\epsilon$ and $\lambda$ of the same order of magnitude,
corrections involving $\epsilon$ produce larger deviations
than corrections involving $\lambda$. This can be seen from the fact
that, in all the cases, the absolute value of the slope of the contour lines
is smaller than one, meaning that, in order to produce modifications
of the same order of magnitude, $|\lambda|$ should be larger than $|\epsilon|$. 
However, if $|\lambda|$ happens to be
much larger than $|\epsilon|$, its effects could naturally
be stronger near the peak of the potential and thus alter
significantly the graybody factor. 
Nonetheless, even for very large $|\lambda|$, as commented above, its
effects would be the fastest to vanish in the asymptotic decay
of the potential.

Plots on the left and right of Fig.~\ref{fig.contours} correspond to small and large
values of the pair $(j,k)$, respectively. In this way one can compare
the effects of increasing these powers. 
The most noticeable effect is that contour lines go from a negative
slope for plots on the left to a positive slope for those on the right.
As already commented above, this shows the change of trend for $\lambda$
depending on small or large values of $k$. That is, for small
values of $k$, the modifications produced by $\lambda$ and $\epsilon$
are correlated, while for large values of $k$, they are anticorrelated.

\subsubsection{Analytic expressions for the graybody factor}\label{sec.graybody}

Let us now turn our attention to the explicit form for the graybody factor
for slightly deformed Schwarzschild black holes.
Using expressions \eqref{eq:gkf}--\eqref{eq:KKon}, and after a linearization in $\epsilon$ and $\lambda$,
we find the following structural form for the graybody factor,
\begin{equation}\label{eq:gkf01}
\sigma_\ell(\nu)=\overset{\circ}\sigma_\ell(\nu)
+\epsilon\,\delta\sigma_\ell^\epsilon(\nu)+\lambda\,\delta\sigma_\ell^\lambda(\nu),
\end{equation}
with $\overset{\circ}\sigma_\ell(\nu)$ being the graybody factor for the Schwarzschild geometry
\begin{equation}
\overset{\circ}\sigma_\ell(\nu)=\frac{1}{1+e^{2\pi \overset{\circ}K_\ell(\nu)}},
\end{equation}
with the exponent function
\begin{equation}\label{eq:Koln}
\overset{\circ}K_\ell(\nu)=\frac{27 (2 \ell (\ell+1)+1) (4 \ell (\ell+1)+3) (11 \ell (\ell+1)+9)-256 (\ell (\ell+1) (62 \ell (\ell+1)+87)+27) \nu ^2}{216 \sqrt{6} (4 \ell (\ell+1)+3))^{5/2}}.
\end{equation}
The correction for the graybody factor contains two pieces at linear order
in the smallness parameters, which read
\begin{align}
\delta\sigma_\ell^\epsilon(\nu)&=2\pi\overset{\circ}\sigma_\ell(\nu)\left(1-\overset{\circ}\sigma_\ell(\nu)\right)E_\ell(\nu)\label{eq:dsige}, \\
\delta\sigma_\ell^\lambda(\nu)&=2\pi\overset{\circ}\sigma_\ell(\nu)\left(1-\overset{\circ}\sigma_\ell(\nu)\right)\Lambda_\ell(\nu),\label{eq:dsigl}
\end{align}
where
the linear correction coefficients $E_\ell(\nu):=-\frac{{\rm d}K_\ell(\nu)}{{\rm d}\epsilon}|_{\epsilon=0}$ and $\Lambda_\ell(\nu):=-\frac{{\rm d}K_\ell(\nu)}{{\rm d}\lambda}|_{\lambda=0}$ are given by
\begin{eqnarray}
    E_\ell(\nu)&=& \frac{-2^{-2 j-\frac{9}{2}}}{27 \sqrt{3} (4 \ell (\ell+1)+3)^{5/2}}(3^j (9 (4 \ell (\ell+1)+3)(2 (8 \ell (\ell+1)(\ell^2+\ell+4)+21) j^2-((4 \ell (\ell+1)+3) j^3)\nonumber\\
    &&-8 \ell
   (\ell+1) (8 \ell (\ell+1)+21) j+12 \ell (\ell+1) (16 \ell (\ell+1)+19)-93 j+54)\nonumber\\
   &&-256 \nu ^2 (-((4 \ell (\ell+1)+3) j^3)+2 (4 \ell (\ell+1) (2\ell (\ell+1)+7)+21) j^2\label{eq:Eln}\\
   &&-8 \ell (\ell+1)
   (4 \ell (\ell+1)+5) j-4 \ell (\ell+1) (32 \ell (\ell+1)+69)-3 (7 j+54)))\nonumber\\
   &&-9( 2^{2 j+1}  (256 (\ell (\ell+1) (6 \ell (\ell+1)+11)+6) \nu^2+9 \ell (\ell+1) (4 \ell (\ell+1)+3) (6
   \ell (\ell+1)+5))),\nonumber\\
   \Lambda_\ell(\nu)&=& \frac{2^{-2 k-\frac{15}{2}} 3^{k-\frac{7}{2}}}{(4 \ell (\ell+1)+3)^{5/2}}(256 \nu ^2 ((4 \ell (\ell+1)+3) k^3-(80 \ell (\ell+1)+69) k^2\nonumber\\
   &&+2 (104 \ell (\ell+1)+123) k+72 (4 \ell (\ell+1)+3))-9 (4 \ell (\ell+1)+3)((4 \ell (\ell+1)+3) k^3\label{eq:Lln}\\
   &&-(88 \ell (\ell+1)+69) k^2+6
   (64 \ell (\ell+1)+53) k-72 (4 \ell (\ell+1)+3))].\nonumber
\end{eqnarray}

\begin{figure}
    \centering
\includegraphics[width=\linewidth]{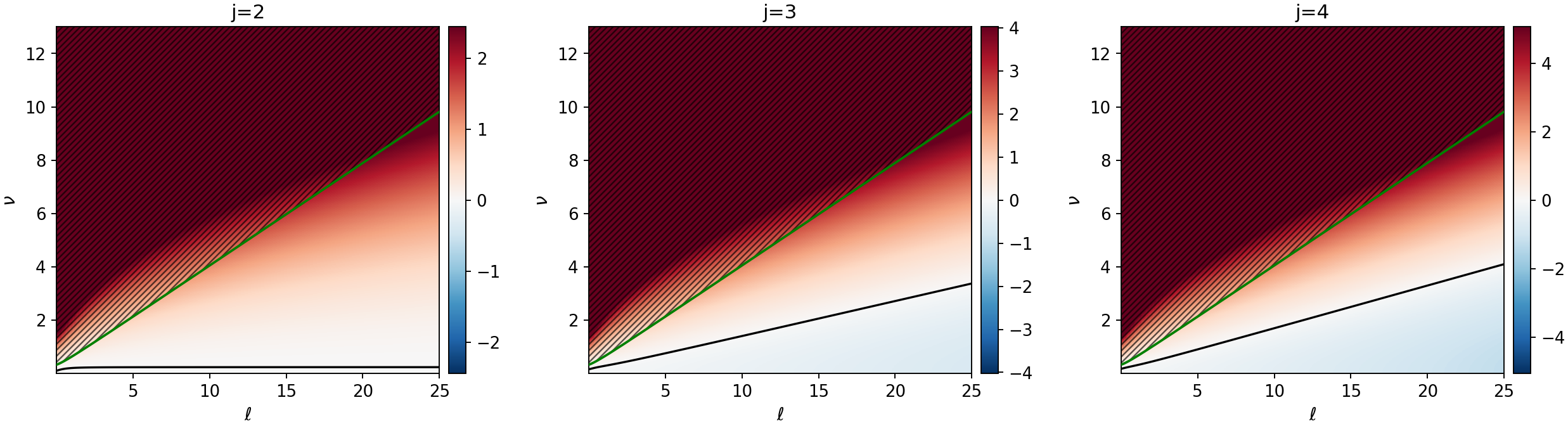}
    \caption{Correction coefficient $E_\ell(\nu)$ over the range $\ell\in[0,25]$, $\nu\in[0,13]$, for the specific values $j=2,3,4$. Red (blue) regions indicate $E_\ell(\nu)>0$ ($E_\ell(\nu)<0$), separated by the solid black curve marking the exact zero-crossing $E_\ell(\nu)=0$. The green curve shows the boundary $\nu^2=\mathring v_\ell(\mathring x_{\max})$, above which the zeroth-order (Schwarzschild) quantity $\mathring d_{\max}^\ell(\nu)$ ceases to be real and thus the approximation is not valid;
    the hatched region marks where this occurs. For all three plotted values of $j$, $E_\ell(\nu)$ is negative
    for low frequencies, becoming  positive only once $\nu$ is sufficiently large relative to $\ell$. Increasing $j$
     enhances the slope of
    the zero-crossing curve.
    }
    \vspace{20pt}
    \label{fig:smE}
\includegraphics[width=0.74\linewidth]{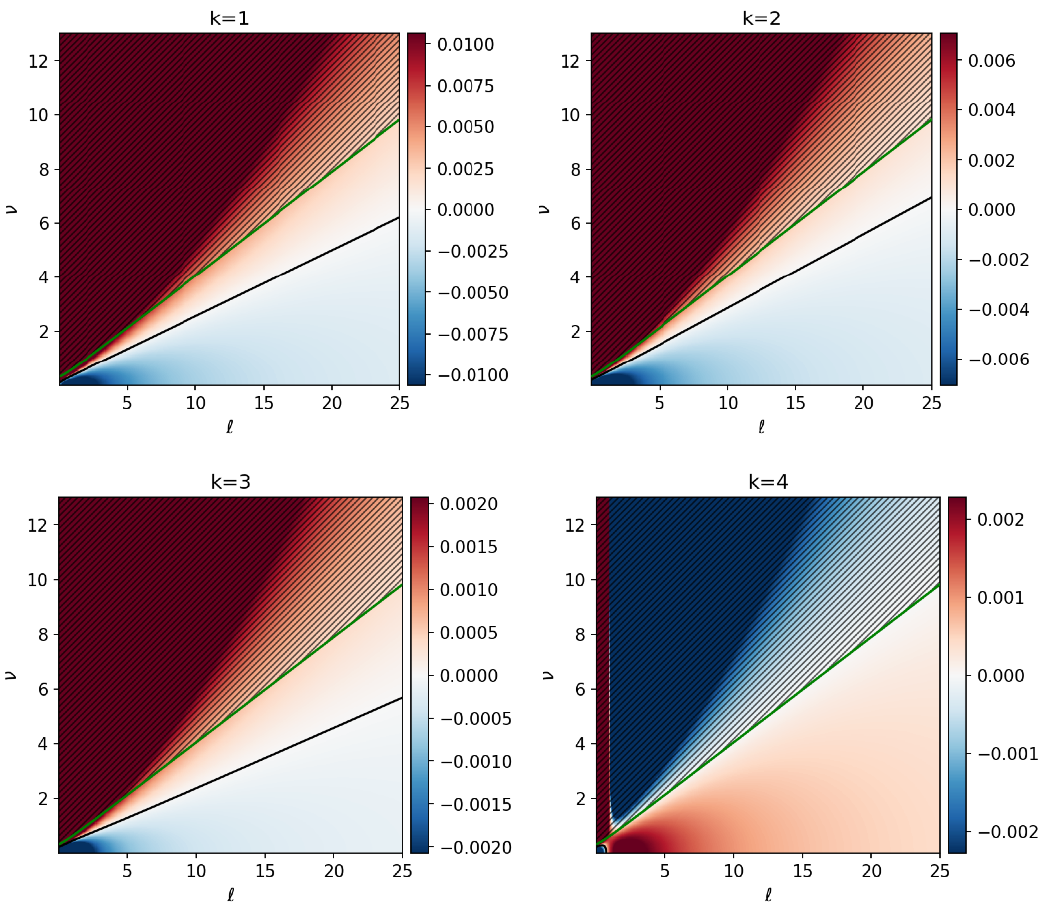}
    \caption{Correction coefficient $\Lambda_\ell(\nu)$, over the range $\ell\in[0,25]$, $\nu\in[0,13]$, for the specific values $k=1,2,3,4$.  As in Fig.~\ref{fig:smE}, red (blue) regions indicate $\Lambda_\ell(\nu)>0$ ($\Lambda_\ell(\nu)<0$), separated by the solid black curve marking the exact zero-crossing $\Lambda_\ell(\nu)=0$ in the allowed region. The hatched region above the green curve $\nu^2=\mathring v_\ell(\mathring x_{\max})$ corresponds to a complex $\mathring d_{\max}^\ell(\nu)$. For $k=1,2,3$, $\Lambda_\ell(\nu)$ shows the same qualitative trend as $E_\ell(\nu)$: it is negative for $\ell$ large relative to $\nu$ and positive once $\nu$ dominates. However, the case $k=4$ is negative close to the origin and becomes positive as we increase either $\ell$ or $\nu$. That is, s-waves ($\ell=0$) have a negative (positive) contribution for low (high) frequencies. But for $\ell\ge 1$, $\Lambda_\ell(\nu)$ is positive for $\ell$ large relative to $\nu$, and negative once $\nu$ dominates. In fact, for $k=4$ the zero-crossing curve approaches the green line asymptotically from above.}
    \label{fig:smL}
\end{figure}

\noindent As expected from the discussion in previous sections,
the modulation to the graybody factor is very cumbersome, and shows a nontrivial
dependence on the mode number $\ell$, the frequency $\nu$, as well as on the powers
of the correction $j$ and $k$.
 However, there are two interesting cases, the s-waves ($\ell=0$)
and the large-$\ell$ limit, which can be analyzed in quite some detail. Both these limits are explicitly studied in App.~\ref{appB}. In summary, on the one hand, for s-waves
we see that $E_0(\nu)$ and $\Lambda_0(\nu)$
do not have a definite sign independent of $j$ and $k$.
However, for small values of $j$ and $k$, both are
typically of the form $-a+b\nu^2$, with positive $a$ and $b$,
while for large values they follow the asymptotics $\lim_{k\to\infty}\Lambda_0(\nu)=0$,
which is independent of $\nu$, while $\lim_{j\to\infty}E_0(\nu)=\frac{32\sqrt2}{27}\nu^2$.
On the other hand,
for large $\ell$ we find that the modifications produced by $\lambda$
and $\epsilon$ follow a different asymptotic behavior: $|E_\ell(\nu)|\sim \ell$,
while $|\Lambda_\ell(\nu)|\sim1/\ell$. Hence, the corrections related to $\epsilon$
grow with $\ell$ similarly as $\overset\circ K_\ell(\nu)$, whereas the corrections related to $\lambda$ decay
for large $\ell$.
Once analyzed those specific limits, let us in the following
extract some qualitative information
about the corrections to the graybody factor for generic $\ell$ modes.

Specifically, one of the relevant points is that, since the global prefactor $\mathring\sigma_\ell(\nu)(1-\mathring\sigma_\ell(\nu))$ in \eqref{eq:dsige} and \eqref{eq:dsigl}
is positive definite, the sign of $\epsilon E_\ell(\nu)$ and $\lambda \Lambda_\ell(\nu)$
completely determines the sign of $\delta\sigma_\ell^\epsilon(\nu)$ and $\delta\sigma_\ell^\lambda(\nu)$,
respectively, and, hence, whether the leading corrections
to the graybody factor increases or decreases
the transmission probability relative to the Schwarzschild value $\mathring\sigma_\ell(\nu)$.
In order to illustrate the sign of these correction terms as a
function of the mode number $\ell$ and the frequency $\nu$, in Figs.~\ref{fig:smE}--\ref{fig:smL}
we present contour plots for $\Lambda_\ell(\nu)$ and $E_\ell(\nu)$ for the first allowed
integer values of $j$ and $k$.

In those contour plots,
red and blue regions correspond to positive and negative signs, respectively,
while the black solid line indicates the zero value of the corresponding quantity ($E_\ell(\nu)$ 
or $\Lambda_\ell(\nu)$).
In addition, as commented above, the present approximation has a validity regime:
for large frequencies $\nu^2> v_\ell(x_{\rm max})$,
for which $\mathring d_{\max}^\ell(\nu)$
becomes imaginary, the approximation cannot be applied and $\sigma_\ell\approx 1$.
In Figs.~\ref{fig:smE}--\ref{fig:smL},
the boundary of the validity region for Schwarzschild, corresponding to $\nu^2 = \mathring v_\ell(\mathring x_{\max})$,
is shown as a green curve, and it follows the simple asymptotic form
$\nu \to \frac{2}{3\sqrt3}\,\ell \approx 0.385\,\ell$ as $\ell\to\infty$.
The region where the approximation is not valid is hatched.

In general, for small values of $j$ and $k$
(represented here by $j=2,3,4$ and $k=1,2,3$),
the contour plots for $E_\ell(\nu)$ and $\Lambda_\ell(\nu)$ are qualitatively
similar. But the behavior may change completely if one considers larger values
of $j$ or $k$, as can be seen in the plot corresponding to $k=4$, which we
show here as a representative example. We refer the reader to App.~\ref{appB} for
more details about large values of $j$ and $k$.

More precisely, for small values of $j$,
as can be seen in the sign plot for $E_\ell(\nu)$ (cf. Fig.~\ref{fig:smE}),
within the validity region, $E_\ell(\nu)$ is negative for low frequencies,
and turns positive once {$\nu$} is sufficiently large compared to $\ell$. The zero-crossing curve $E_\ell(\nu)=0$ approaches
 the straight line $\nu \to c_E(j)\,\ell$ at large $\ell$, with $c_E(2)\approx0$, $c_E(3)\approx0.132$, and $c_E(4)\approx0.161$---consistently smaller than the Schwarzschild validity slope, which is $0.385$.
Thus,
increasing $j$ from 2 to 4 enhances the slope of the zero-crossing curve for $\ell\gtrsim 1$,
and higher-$j$ corrections remain negative over a larger portion of the validity region in the $(\ell,\nu)$ plane.
As already commented, for small $k$ (here represented by $k=1,2,3$),
the sign plot for
$\Lambda_\ell(\nu)$ (cf. Fig.~\ref{fig:smL}) shows the same trend as $E_\ell(\nu)$:
it is negative for $\ell$ large relative to $\nu$, and positive once $\nu$ dominates, crossing zero along $\nu\to c_\Lambda(k)\,\ell$
with $c_\Lambda(1)\approx0.243$, $c_\Lambda(2)\approx0.272$, and $c_\Lambda(3)\approx0.222$.
However, in this case the black solid line does not approach the green line
monotonically with an increasing $k$.

\begin{figure}
    \centering
\includegraphics[width=0.55\linewidth]{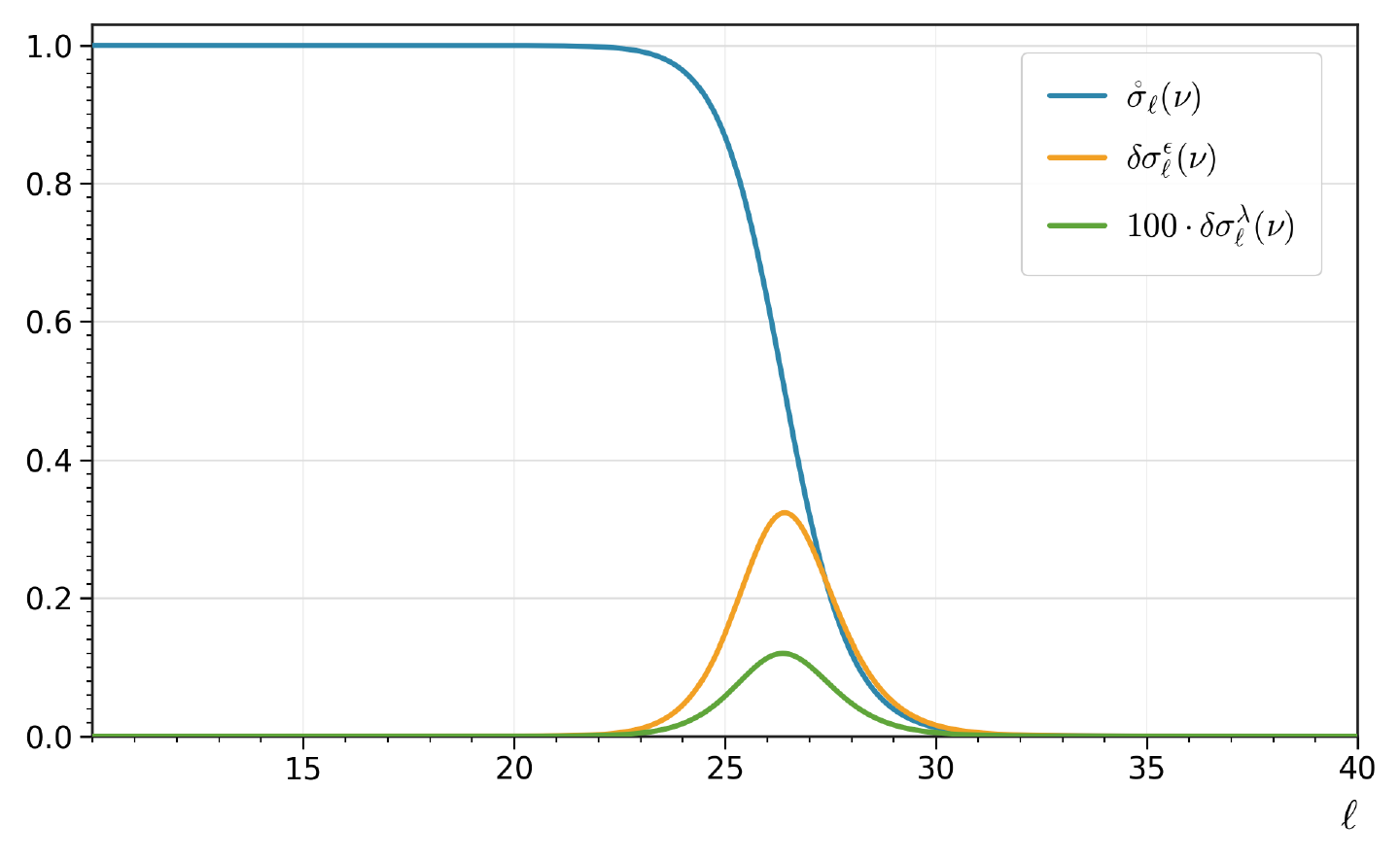}
    \caption{
    The blue curve corresponds to the graybody factor for the Schwarzschild black hole $\overset{\circ}\sigma_l$ as a function of $\ell$,
    and for a fixed frequency $\nu=0.1$. For small $\ell$ the reflection is almost negligible $(\overset{\circ}\sigma_\ell\approx 1)$.
    At about $\ell\approx 23$ the scattering ramps up, and the potential becomes more reflective, leading to a large
    scattering barrier. For very large $\ell$, the potential barrier blocks completely the
    radiation, and thus $\overset{\circ}\sigma_\ell(\nu)\approx 0$. The corrections are displayed individually and evaluated with the same frequency $\nu=0.1$.
    The orange curve shows the $\epsilon$-related correction $\delta\sigma_\ell^\epsilon$ for $j=2$.
    The green curve depicts $\delta\sigma_\ell^\lambda$ with $k=3$, but, since this contribution is much smaller than $\delta\sigma_\ell^\epsilon$,
    to be able to show them in the same plot, we multiplied it by a factor of 100.
        }
    \label{fig:figure5}
\end{figure}

As explained above, the sign of the correction to the graybody factor
is completely encoded in the combinations $\epsilon E_{\ell}$ and
$\lambda\Lambda_{\ell}$, and, for a given sign of $\lambda$ and $\epsilon$,
one can immediately read the sign of the individual correction terms
from the plots discussed above.
However, the magnitude of the modification is tuned by the global factor
$\overset{\circ}{\sigma}_\ell(\nu)(1-\overset{\circ}{\sigma}_\ell(\nu))$,
which appears in the complete expression of the corrections \eqref{eq:dsige}--\eqref{eq:dsigl}. In order to see its effects
for a simple showcase example,
in Fig.~\ref{fig:figure5} we present the different contributions to the graybody factor as a function of the mode
number $\ell$, for the values $j=2$, $k=3$, and $\nu=0.1$.
The Schwarzschild graybody factor $\overset{\circ}{\sigma}_\ell(\nu)$ is plotted in blue,
while the individual corrections $\delta\sigma_\ell^\epsilon(\nu)$ and $\delta\sigma_\ell^\lambda(\nu)$
are displayed in orange and green, respectively.
Since the height of the potential increases with $\ell$, the Schwarzschild graybody factor
$\overset{\circ}{\sigma}_\ell(\nu)$ goes from 1 for small $\ell$ to 0 for large $\ell$,
and shows an inflection point around $\ell\approx 23$.
This behavior of $\overset{\circ}{\sigma}_\ell(\nu)$ makes the global factor
$\overset{\circ}{\sigma}_\ell(\nu)(1-\overset{\circ}{\sigma}_\ell(\nu))$
to be vanishing for both small and large $\ell$.
In consequence, despite the magnitude of $E_\ell$ and $\Lambda_\ell$
in different parts of the $(\ell,\nu)$ plane,
the support of $\delta\sigma_\ell^\epsilon(\nu)$ and
$\delta\sigma_\ell^\lambda(\nu)$ is mainly confined around the inflection point.

\begin{figure}
    \centering
\includegraphics[width=\linewidth]{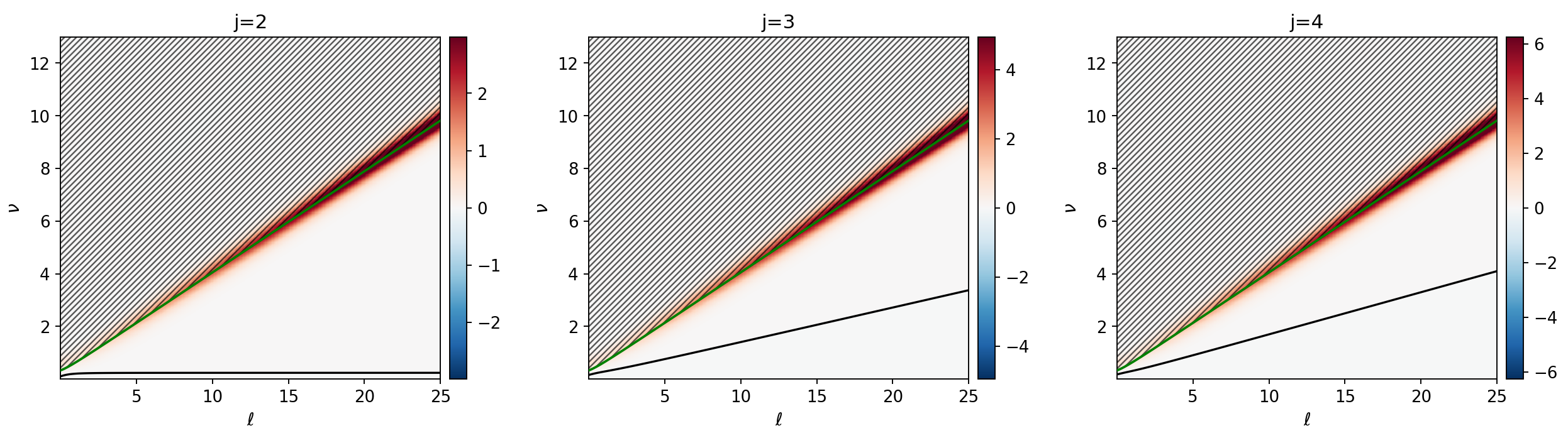}
    \caption{Correction of the graybody factor $\delta\sigma^\epsilon_\ell(\nu)$ over the range $\ell\in[0,25]$, $\nu\in[0,13]$, for the specific values $j=2,3,4$. Red (blue) regions indicate $\delta\sigma^\epsilon_\ell(\nu)>0$ ($\delta\sigma^\epsilon_\ell(\nu)<0$), separated by the solid black curve marking the exact zero-crossing $\delta\sigma^\epsilon_\ell(\nu)=0$. The green curve shows the boundary $\nu^2=\mathring v_\ell(\mathring x_{\max})$, above which the zeroth-order (Schwarzschild) quantity $\mathring d_{\max}^\ell(\nu)$ ceases to be real; the hatched region marks where this occurs. For all three plotted values of $j$, $\delta\sigma^\epsilon_\ell(\nu)$ is mostly positive and concentrated around the green curve.
These plots show explicitly the correction to the graybody factor corresponding to specific black-hole geometries:
Reissner-Nordstr\"om $(j=2)$, Bardeen $(j=3)$, and Hayward $(j=4)$.
}
    \label{fig:gkfsigmaepsw}
    \vspace{20pt}
\includegraphics[width=0.74\linewidth]{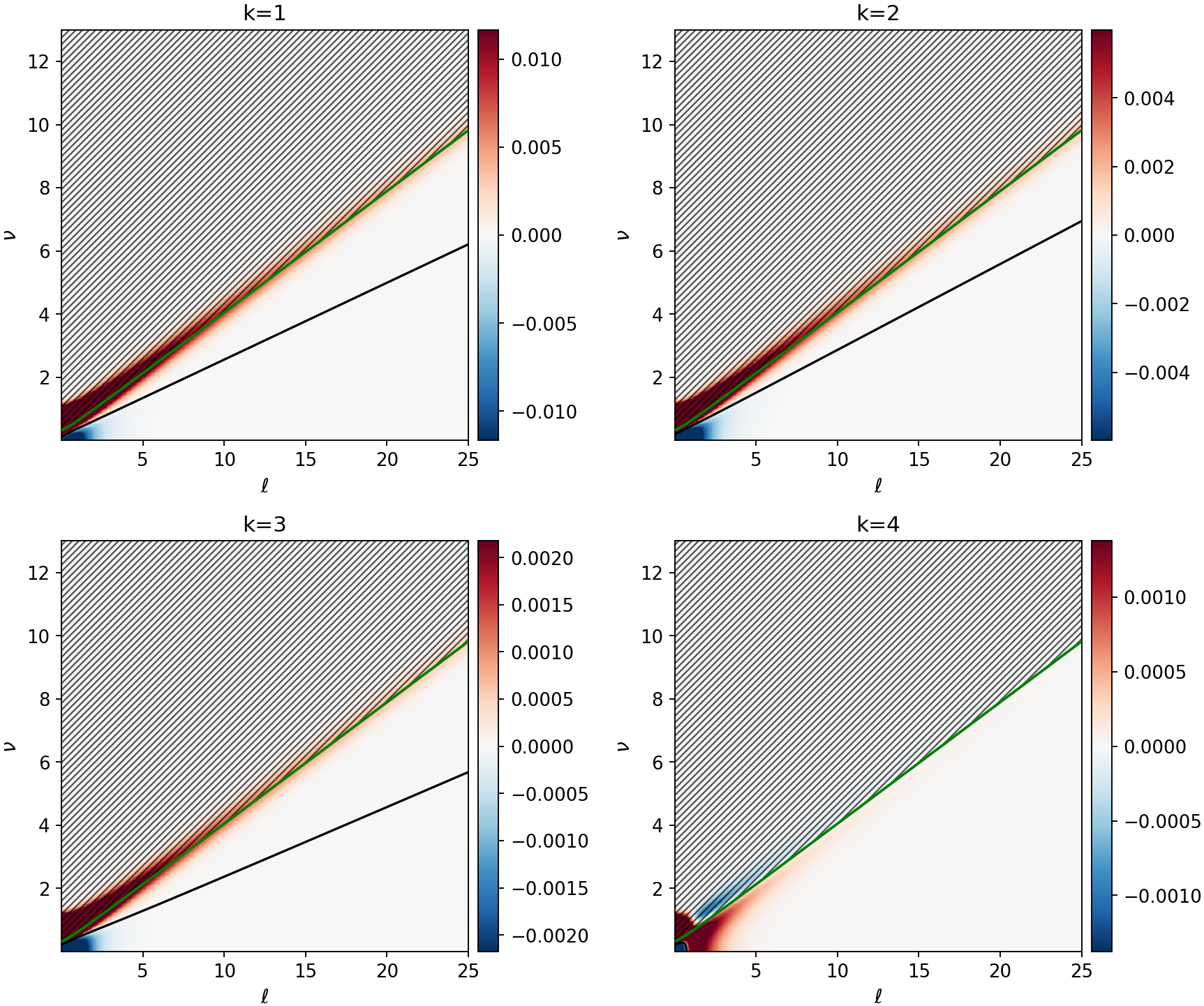}
    \caption{Correction of the graybody factor $\delta\sigma^\lambda_\ell(\nu)$  over the range $\ell\in[0,25]$, $\nu\in[0,13]$, for the specific values $k=1,2,3,4$. The hatched region indicates where the approximation breaks down. As in Fig.~\ref{fig:gkfsigmaepsw}, red (blue) regions indicate $\delta\sigma^\lambda_\ell(\nu)>0$ ($\delta\sigma^\lambda_\ell(\nu)<0$), separated by the solid black curve marking the exact zero-crossing $\delta\sigma^\lambda_\ell(\nu)=0$ in the allowed region.
    In the first three $(k=1,2,3)$, 
the correction $\delta\sigma^\lambda_\ell(\nu)$ is mostly concentrated in two regions:
around the green curve, where $\delta\sigma^\lambda_\ell(\nu) >0$,
and close to the origin, where $\delta\sigma^\lambda_\ell(\nu)<0$. In the last case ($k=4)$, while we keep finding $\delta\sigma^\lambda_\ell(\nu)<0$ close to the origin (for $\ell=0$ and small $\nu$), it becomes positive in the validity region as soon as we
increase a bit $\nu$ or take $\ell\geq 1$.  
These plots show explicitly the correction to the graybody factor corresponding to specific black-hole geometries:
$r_0$-model $(k=1)$ and Simpson-Visser $(k=2)$. Note also that the magnitude of the correction diminishes as we increase $k$.
}
    \label{fig:gkfsigmalamw}
\end{figure}
\begin{figure}
    \centering
\includegraphics[width=\linewidth]{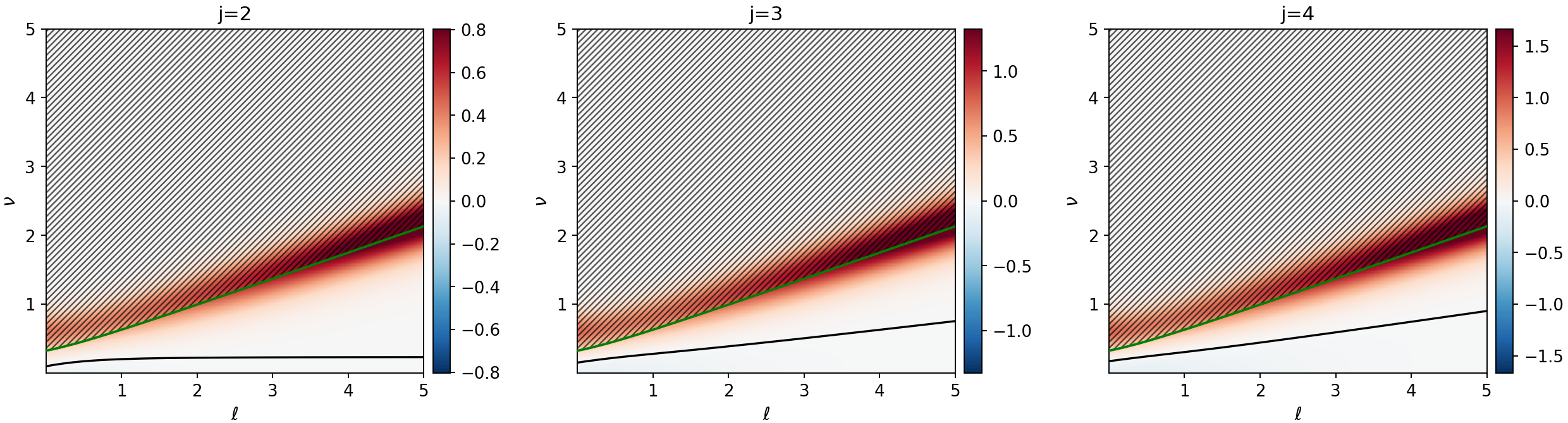}
    \caption{Zoom in of Fig.~\ref{fig:gkfsigmaepsw}, representing the correction of the graybody factor $\delta\sigma^\epsilon_\ell(\nu)$ over the range $\ell\in[0,5]$, $\nu\in[0,5]$, for the specific values $j=2,3,4$.
}
    \label{fig:gkfsigmaeps}
    \vspace{20pt}
\includegraphics[width=0.74\linewidth]{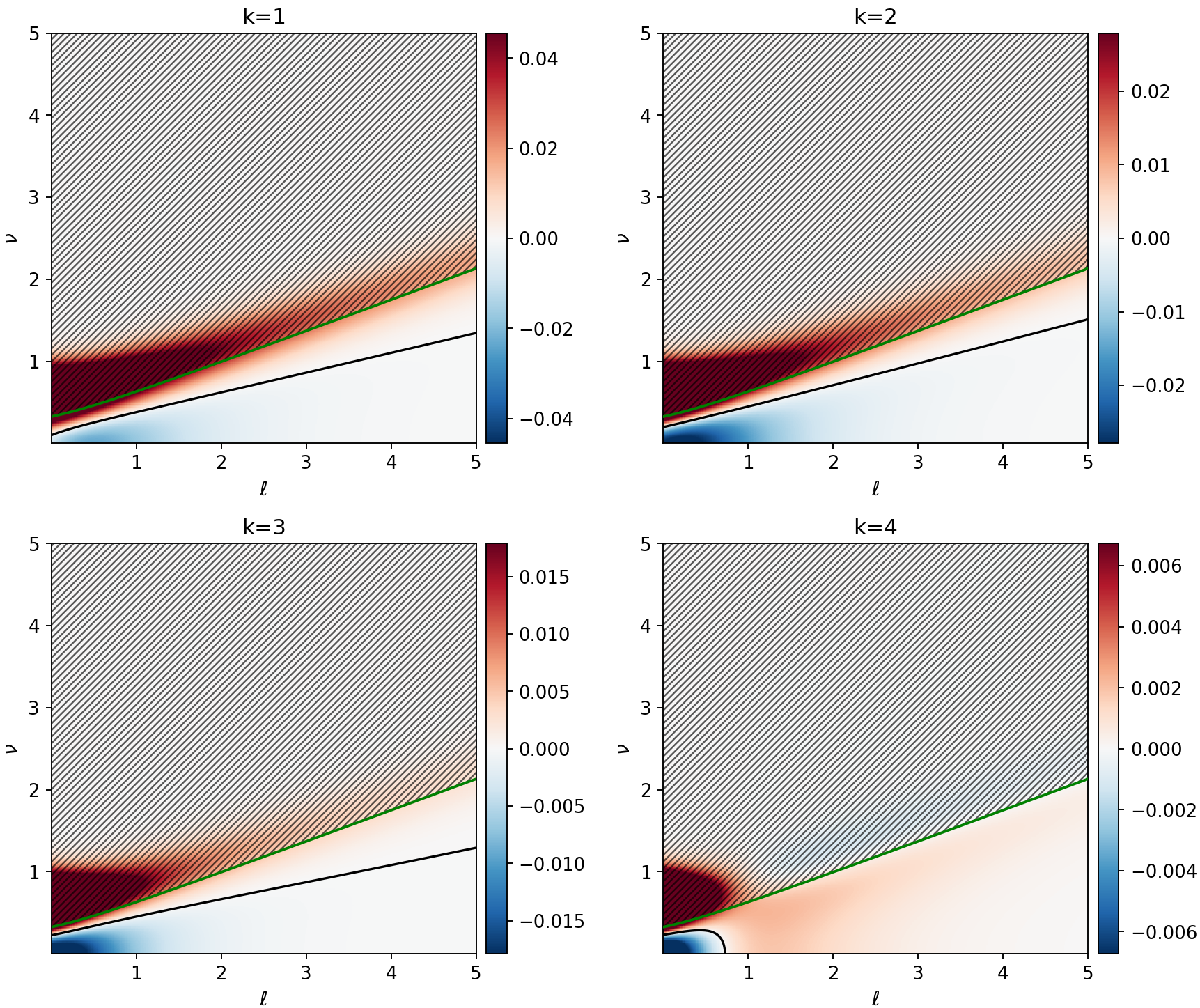}
    \caption{Zoom in of Fig.~\ref{fig:gkfsigmalamw}, representing the correction of the graybody factor $\delta\sigma^\lambda_\ell(\nu)$ over the range $\ell\in[0,5]$, $\nu\in[0,5]$, for the specific values $k=1,2,3,4$.
}   
   \label{fig:gkfsigmalam}
\end{figure}

This turns out to be a very general property of the corrections to the graybody
factor, as can be seen in Figs.~\ref{fig:gkfsigmaepsw}--\ref{fig:gkfsigmalam}.
These figures show contour plots of the full corrections,
$\delta\sigma_\ell^\epsilon(\nu)$ and $\delta\sigma_\ell^\lambda(\nu)$,
in different regions of the $(\ell,\nu)$ plane for fixed values of $j$ and $k$.
Thus, even though $E_\ell(\nu)$ and $\Lambda_\ell(\nu)$ may become large away
from the inflection point (which in these plots correspond to the green line), the prefactor induces an exponential suppression, such that the relevant part of the correction is around the green line. 
For small values of $j$ and $k$, plots for $\delta\sigma_\ell^\epsilon(\nu)$
and $\delta\sigma_\ell^\lambda(\nu)$ seem qualitatively similar.
The main difference is that,
in addition to the region around the green line, $\delta\sigma_\ell^\lambda(\nu)$
also shows large values around the origin of the $(\ell, \nu)$ plane.
Also, the maximum value of $\delta\sigma_\ell^\epsilon(\nu)$
is typically 2-3 orders of magnitude larger than the value corresponding to $\delta\sigma_\ell^\lambda(\nu)$.
The features shown in these plots are of key relevance since,
as will be commented in the next section, some of the considered values of the parameters $j$ and $k$
correspond to specific black-hole geometries.

\section{Application to particular black-hole spacetimes}\label{sec:apltpbhsts}

To understand how the corrections affect the radiation emitted by specific black-hole geometries,
we calculate the temperatures, and comment on the modifications to the Schwarzschild graybody factor,
for the following particular spacetimes: the Reissner-Nordstr\"on, the Bardeen, the Hayward, and
the Simpson-Visser spacetimes, as well as some regular black-hole models inspired by quantum gravity.

\paragraph{Reissner-Nordström:}
The Reissner-Nordström metric,
\begin{align}\label{eq.rn}
  g_Q=-\left(1-\frac{2M}{r}+\frac{Q^2}{r^2}\right)\mbox{d}t\otimes\mbox{d}t+\left(1-\frac{2M}{r}+\frac{Q^2}{r^2}\right)^{-1}\mbox{d}r\otimes\mbox{d}r+r^2\mbox{d}\mathbb{S}_2,
  \end{align}
is a solution of the Einstein equations and describes a black hole with electric charge $Q$.
As Schwarzschild, this spacetime admits still a singularity at $r=0$, although the singular structure is timelike.
Comparing this line element with our general parameterization \eqref{eq:metrikG-param}, and with the definitions
\eqref{defG} and \eqref{defh3} for the
shape functions, it is easy to see that, for $Q\ll M$, it fits the scheme of slightly deformed Schwarzschild black
holes with parameters $\epsilon=Q^2/(2M)^2$, $\lambda=0$, and $j=2$.

\paragraph{Bardeen:} The Bardeen spacetime is one of the pioneering models of a regular black hole. It is constructed such that the metric behaves like a de Sitter space
near $r=0$, where curvature invariants remain finite. The metric is characterized by the mass parameter $M$ and a magnetic charge parameter $q$, which modifies the standard Schwarzschild geometry
\cite{bardeen},
\begin{align}\label{eq.bardeenm}
    g_q&=-\left(1-\frac{2Mr^2}{(q^2+r^2)^{3/2}}\right)\mbox{d}t\otimes\mbox{d}t+\left(1-\frac{2Mr^2}{(q^2+r^2)^{3/2}}\right)^{-1}\mbox{d}r\otimes\mbox{d}r+r^2\mbox{d}\mathbb{S}_2.
\end{align}
Performing a Taylor expansion yields
\begin{align*}
    g_q&=-\left(1-\frac{2M}{r}+\frac{3Mq^2}{r^3}+{\cal O}\left(\frac{Mq^4}{r^5}\right)\right)\mbox{d}t\otimes\mbox{d}t+\left(1-\frac{2M}{r}+\frac{3Mq^2}{r^3}+{\cal O}\left(\frac{Mq^4}{r^5}\right)
    \right)^{-1}\mbox{d}r\otimes\mbox{d}r+r^2\mbox{d}\mathbb{S}_2.
\end{align*}
Therefore, assuming that $|q|\ll 2 M$, outside the horizon
this geometry is accurately described by
\begin{align*}
    g_q
    &\approx-\left(1-\frac{2M}{r}+\frac{3Mq^2}{r^3}\right)\mbox{d}t\otimes\mbox{d}t+\left(1-\frac{2M}{r}+\frac{3Mq^2}{r^3}\right)^{-1}\mbox{d}r\otimes\mbox{d}r+r^2\mbox{d}\mathbb{S}_2.
\end{align*}
From here we can directly read the shape parameters for the Bardeen black hole: $\epsilon=\frac{3q^2}{8 M^2}$, $\lambda=0$, and $j=3$.

\paragraph{Hayward:} The Hayward metric asymptotes to Schwarzschild
for large values of the areal radius function,
and transitions to a de Sitter core in the interior of the black hole \cite{Hayward:2005gi}.
This model captures a localized region of negative pressure at the center,
preventing gravitational collapse from reaching smaller volumes than $\beta>0$. Its line element reads,
\begin{align}
    g_\beta&=-\left(1-\frac{2Mr^2}{r^3+\beta}\right)\mbox{d}t\otimes\mbox{d}t+\left(1-\frac{2Mr^2}{r^3+\beta}\right)^{-1}\mbox{d}r\otimes\mbox{d}r+r^2\mbox{d}\mathbb{S}_2.
\end{align}
Following the same steps as for the Bardeen black hole, and assuming now
that $\beta\ll (2 M)^3$, outside the horizon
the Hayward geometry can be well approximated by the line element,
\begin{align*}
g_\beta\approx
-\left(1-\frac{2M}{r}+\frac{2M\beta}{r^4}\right)\mbox{d}t\otimes\mbox{d}t+\left(1-\frac{2M}{r}+\frac{2M\beta}{r^4}\right)^{-1}\mbox{d}r\otimes\mbox{d}r+r^2\mbox{d}\mathbb{S}_2,
\end{align*}
and, hence, it is described through the parameters $\epsilon=\beta/(2 M)^3$, $\lambda=0$, and $j=4$ in our ansatz.
It is interesting to note that the black-hole geometry presented in Ref.~\cite{Kelly:2020uwj}, with line element,
\begin{align*}
   g_\alpha=-\left(1-\frac{2M}{r}+\frac{\alpha}{r^4}\right)\mbox{d}t\otimes\mbox{d}t+\left(1-\frac{2M}{r}+\frac{\alpha}{r^4}\right)^{-1}\mbox{d}r\otimes\mbox{d}r+r^2\mbox{d}\mathbb{S}_2,
\end{align*}
for certain $\alpha>0$, is described by the same set of parameters
$\epsilon=\alpha$, $\lambda=0$, and $j=4$ as the (approximated) Hayward geometry. 
Thus, even if the interior of the horizon could feature very different properties,
the only form of the line element that matters for the study of the temperature and the graybody factor is that outside the horizon. Therefore, this geometry is included in our
discussion of the Hayward metric by simply taking $\beta=\alpha/(2M)$.

\paragraph{Simpson-Visser:} The Simpson-Visser metric provides a one-parameter generalization of Schwarzschild that represents black holes, as well as traversable wormholes, depending on the ratio of the mass $M$ and the additional parameter $a$ \cite{Simpson:2018tsi}. Thus, the Simpson-Visser spacetime can represent a regular black hole with two horizons, an extremal black hole, a one-way horizonless throat, or a traversable wormhole. Its line element
is given by
\begin{align}
    g_a=-\left(1-\frac{2M}{\sqrt{r^2+a^2}}\right)\mbox{d}t\otimes\mbox{d}t+
    \left(1-\frac{2M}{\sqrt{r^2+a^2}}\right)^{-1} \mbox{d}r\otimes\mbox{d}r+(r^2+a^2) \mbox{d}\mathbb{S}_2.
\end{align}
It is straightforward to check that, a shift of the area radius function $r^2\to(r^2-a^2)$ yields the form
\begin{align*}
    g_a=-\left(1-\frac{2M}{r}\right)\mbox{d}t\otimes\mbox{d}t+
    \left(1-\frac{2M}{r}\right)^{-1} \left(1-\frac{a^2}{r^2}\right)^{-1}\mbox{d}r\otimes\mbox{d}r+r^2 \mbox{d}\mathbb{S}_2,
\end{align*}
 from where we can read its parameters $\epsilon=0$, $\lambda=a^2/(2 M)^2$, and $k=2$.

\paragraph{$r_0$-model:} This spacetime represents a deviation from the Schwarzschild black hole inspired by loop quantum gravity.
The singularity is here removed because the future trapped region ends in a minimal surface foliated by spheres of minimum area $4\pi r_0^2<16\pi M^2$, which
opens up into a white-hole region \cite{Alonso-Bardaji:2021yls,Alonso-Bardaji:2022ear}.
This geometry is described by the following line element,
\begin{align}\label{eq.ourmetric}
    g_{r_0}&=-\left(1-\frac{2M}{r}\right)\mbox{d}t\otimes\mbox{d}t+\bigg(1-\frac{r_0}{r}\bigg)^{-1}\left(1-\frac{2M}{r}\right)^{-1}\mbox{d}r\otimes\mbox{d}r+r^2 \mbox{d}\mathbb{S}_2,
\end{align}
 which defines the set of parameters $\epsilon=0$, $\lambda=r_0/(2 M)$, and $k=1$.

\paragraph{$\Delta$-model:} An alternative black-hole geometry, constructed also in the context of loop quantum gravity,
is described by \cite{Belfaqih:2024vfk}
 \begin{align}\label{eq:deltametrik}
     g_{\Delta}&=-\left(1-\frac{2M}{r}\right)\mbox{d}t\otimes\mbox{d}t+\bigg(1+\frac{\Delta}{r^2}\left(1-\frac{2M}{r}\right)\bigg)^{-1}\left(1-\frac{2M}{r}\right)^{-1}\mbox{d}r\otimes\mbox{d}r+r^2 \mbox{d}\mathbb{S}_2.
 \end{align}
This line element
is included among the family \eqref{eq:metrikh123}, with unmodified shape functions
$h_1(r)=1$ and $h_2(r)=r$ with respect to Schwarzschild, but with
$h_3(r)$ determined by two terms with different powers of $r$.
Therefore, it does not fit our ansatz \eqref{defh3}
for slightly deformed Schwarzschild geometries. However, the application
of the formalism to this model is straightforward, and, for completeness,
in App.~\ref{appA} we present in detail the derivation of its corresponding graybody factor, which complements the study
performed in Ref.~\cite{Belfaqih:2026qgj} mainly focused on s-waves $(\ell=0)$.

Overall, the above geometries can be split in two kinds.
On the one hand, the Reissner-Nordstr\"on, Bardeen, and Hayward
geometries modify the norm of the Killing field. Thus the position of the horizon differs
from its Schwarzschild counterpart, as it is given by $r_H=2M(1-\epsilon)$ to leading order
in $\epsilon$, and corrections correspond to $\epsilon$ modifications.
On the other hand, the Simpson-Visser, the $r_0$-model, and the $\Delta$-model do not introduce any
corrections in the Killing norm, as they only carry modifications of the $h_3$ shape function,
which are parameterized by $\lambda$. 

Let us now discuss the temperature of the horizon defined by the above
geometries, which directly follows from 
\eqref{eq:temperatur}.
Here we present the linearized  correction
$\delta T_H$, as given by \eqref{eq.linearizedTcorrection},
for each corrected spacetime:
\begin{eqnarray}
 \mbox{\textit{Reissner-Nordström}}&:& \delta  T_H^Q=-\frac{\hbar Q^2}{4\pi r_H^3}+\mathcal{O}(Q^4),\\
 \mbox{\textit{Bardeen}}&:& \delta T_H^q=-\frac{3\hbar q^2}{4\pi r_H^3}+\mathcal{O}(q^4),\\
  \mbox{\textit{Hayward}}&:& \delta  T_H^\beta=-\frac{3\hbar\beta}{4\pi r_H^4}+\mathcal{O}(\beta^\frac{3}{2}),\\
     \mbox{\textit{Simpson-Visser}}&:&\delta T_H^a=-\frac{\hbar a^2}{8\pi r_H^3}+\mathcal{O}(a^4),\\
   \mbox{$r_0$\textit{-model}}&:& \delta T_H^{r_0}=-\frac{\hbar r_0}{8\pi r_H^2}+\mathcal{O}(r_0^2).
\end{eqnarray}
By comparing the above calculated values, we find an interesting pattern:
all these corrections come with a reduction of the temperature compared to the Schwarzschild case.
Therefore, we have that all the geometries \eqref{eq.rn}--\eqref{eq.ourmetric}
define a colder horizon than
its Schwarzschild counterpart. In particular, this implies a reduced radiation
due to the Hawking effect.
 Still, we have the exception of \eqref{eq:deltametrik}, for which the correction on the horizon exactly vanishes, and thus $\delta T_H^{\Delta}=0$. 

Concerning the corrections to the graybody factor, as explained above,
Reissner-Nordstr\"on, Bardeen, and Hayward are described by positive $\epsilon$ corrections with $j=2$, $j=3$,
and $j=4$ respectively. Therefore, their corresponding correction to the Schwarzschild graybody factor
$\epsilon\,\delta\sigma_{\ell}^\epsilon$
can be seen in the left, middle, and right panel of Fig.~\ref{fig:gkfsigmaepsw}, respectively,
with corresponding zoomed-in versions in Fig.~\ref{fig:gkfsigmaeps}.
On the other hand, the $r_0$-model and Simpson-Visser present no
$\epsilon$ corrections and their corresponding $\lambda$ is positive,
thus their modification to the graybody factor $\lambda\,\delta\sigma_{\ell}^\lambda$ can be explicitly seen
in the  first $(k=1)$ and second $(k=2)$ plots of Fig.~\ref{fig:gkfsigmalamw}, respectively,
with corresponding zoomed-in versions in Fig.~\ref{fig:gkfsigmalam}.
The $\Delta$-model is analyzed in its exact form in App.~\ref{appA}, where we see that its corresponding modification to the
Schwarzschild graybody factor can be approximately described
by the values $\epsilon=0$, $\lambda\approx-\Delta/(2M)^2$, and $k\approx3$,
which correspond to the third plot in Figs.~\ref{fig:gkfsigmalamw} and \ref{fig:gkfsigmalam}.
However, note that, since for this model $\lambda<0$, the full correction $\lambda\,\delta\sigma^\lambda_\ell$
presents an inverted sign with respect to the plotted $\delta\sigma^\lambda_\ell$.

It is very interesting to see that the correction to the graybody factor produced by the different spacetimes \eqref{eq.rn}--\eqref{eq.ourmetric} 
shows the same qualitative behavior.
More precisely, as can be seen in the plots in Figs.~\ref{fig:gkfsigmaepsw}--\ref{fig:gkfsigmalam} corresponding
to the values of $j$ and $k$ of these black-hole models,
the largest values of the corrections mainly concentrate
close to the green curve, which corresponds to
frequencies near the maximum of the potential, $\nu^2\approx v_\ell(x_{\rm max})$.
In addition to this localization, in the plots describing $\lambda$
corrections (corresponding to the Simpson-Visser and $r_0$-model)
we also observe large values of the corrections around the origin.
Interestingly, regarding the sign of the
correction, for all these spacetimes, one can split the frequencies into two types:
\begin{itemize}
\item For frequencies of the order of the height of the potential and a bit lower,
$\nu^2\lesssim v_\ell(x_{\max})$, one finds a positive correction,
and thus an enhanced value of the graybody factor as compared to its Schwarzschild counterpart,
allowing more outgoing modes to be transmitted.
\item For relatively lower frequencies, the correction turns out to
be negative, and thus it reduces the value of the graybody factor, which means more reflection of the outgoing modes.
\end{itemize}
The limit between the two cases is given by the black solid line in Figs.~\ref{fig:gkfsigmaepsw}--\ref{fig:gkfsigmalam},
which corresponds to an exactly vanishing value of the correction.
These properties can be traced back to the shape of the potential. More precisely,
this comes from the fact that, as can be seen in the plots on the left of Figs.~\ref{fig:contour_vmax}--\ref{fig:contour_width},
corrections with positive $\epsilon$ and $\lambda$ and small values of $j$
and $k$, as it is the case for the analyzed black-hole models \eqref{eq.rn}--\eqref{eq.ourmetric},
produce a lower and wider potential as compared to the Schwarzschild geometry.

Notice, however, that the modifications induced by the geometry \eqref{eq:deltametrik}
exhibit just the opposite trend to the other geometries discussed above, yielding a larger
graybody factor at low frequencies and a smaller one at high frequencies.
The same behavior would arise for any spacetime with negative correction parameters $\epsilon$ or $\lambda$.

In summary, we have found that, with the exception of the $\Delta$-model, all the above black-hole geometries
induce similar qualitative changes in the radiative properties of the horizon as compared to Schwarzschild.
On the one hand, they
contain
a colder horizon than their Schwarzschild counterpart. On the other hand, concerning the graybody
factor, for relatively high frequencies, $\nu^2\lesssim v_\ell(\mathring x_{\max})$,
the corrections allow more outgoing modes to tunnel, making the potential
more transparent. However, outgoing modes with relatively lower frequencies,
perceive an effectively more difficult barrier to tunnel through,
and their spectrum is suppressed.

\section{Conclusions}\label{sec.concl}

We have studied the radiative properties of a broad family of static and spherically symmetric black holes~\eqref{eq:metrikh123} described by the three shape functions
$h_1(r), h_2(r),$ and $h_3(r)$.
This family encompasses a variety of regular and quantum-corrected black-hole geometries, and allows us to investigate the relation between
geometric modifications and the Hawking radiation in a unified way. Since different notions of black-hole mass need not coincide for the geometries
under consideration (as presented in Sec.~\ref{sec.mass}), we have compared black holes of equal size, defined by the same horizon area $4\pi r_H^2$.
In principle, this is an observational quantity, which is geometrically well defined.

As it is well known, the dynamics of a minimally coupled scalar field can be reduced to a two-dimensional wave equation \eqref{eq:RWG} on a flat background, with
the Regge-Wheeler potential \eqref{potential} encoding the curvature. This potential naturally separates into a centrifugal
contribution, which depends on the angular mode, and an additional
curvature contribution determined by the radial geometry. For the family of geometries considered here, the potential vanishes both at
the horizon and at spatial infinity and, under the mild monotonicity assumptions assumed in the paper, it exhibits the characteristic
hilltop-shape barrier structure.

We have then derived the Hawking temperature directly from the near-horizon behavior of the metric. The result \eqref{eq:temperatur} depends on the derivative
of the norm of the Killing vector at the horizon, with the shape function $h_3$ entering as a global multiplicative factor evaluated at the horizon.
This illustrates the local character of the temperature, in contrast with the graybody factor, which depends on the scattering properties
of the whole exterior geometry. In particular, as shown in Eqs. \eqref{eq:gkf}--\eqref{eq:KKon},
in a first approximation the graybody factor is determined by the height and curvature of the potential around the maximum.

Subsequently, we considered small deformations around the Schwarzschild geometry, introducing two deformation parameters,
$\epsilon$ and $\lambda$, together with their corresponding radial powers $j$ and $k$. At linear order in $\epsilon$ and $\lambda$,
the correction to the Schwarzschild temperature \eqref{eq.linearizedTcorrection} takes a particularly simple form.
For fixed horizon radius,
positive values of either deformation parameters lower the temperature, whereas negative values increase it.
Interestingly, the power $k$ governing the radial dependence of the $h_3$ deformation does not enter the
temperature.

The corrections to the graybody factor are considerably richer. They depend not only on the deformation parameters
and their radial powers, but also on the frequency $\nu$ and mode number $\ell$. This dependence originates
from the modification of the height, width, and location of the Regge-Wheeler potential barrier. In particular,
for $\epsilon$ and $\lambda$ of comparable magnitude,
deformations associated with $\epsilon$ are typically more important than those associated with $\lambda$. 
In addition, we observe that in general
the graybody-factor corrections are strongly localized in two regions
of the $(\ell,\nu)$ plane:
around the region where the Schwarzschild transmission coefficient changes from nearly transparent
to nearly opaque (near the green line of Figs~\ref{fig:gkfsigmaepsw}--\ref{fig:gkfsigmalam}), and around the origin (small $\ell$ and $\nu$).

Finally, we have applied the framework to several specific black holes,
including Reissner-Nordström, Bardeen, Hayward, Simpson-Visser,
and certain geometries inspired by loop quantum gravity.
To leading order, with the exception of the $\Delta$-model \eqref{eq:deltametrik} that features no corrections in its temperature,
all the considered black holes present a lower temperature than a Schwarzschild black hole of the same horizon size.

Since the corresponding correction parameter,
$\epsilon$ or $\lambda$, is positive for all the models \eqref{eq.rn}--\eqref{eq.ourmetric}, the full
corrections to the graybody factor
for the different geometries, that is, $\epsilon\delta\sigma^\epsilon_\ell$ or $\lambda \delta\sigma^\lambda_\ell$, can be directly inferred from  Figs.~\ref{fig:gkfsigmaepsw}--\ref{fig:gkfsigmalam}, which correspond to:
Reissner-Nordstr\"om (left plot on Figs.~\ref{fig:gkfsigmaepsw} and \ref{fig:gkfsigmaeps}), Bardeen (middle plot on Figs.~\ref{fig:gkfsigmaepsw} and \ref{fig:gkfsigmaeps}),
Hayward (right plot on Figs.~\ref{fig:gkfsigmaepsw} and \ref{fig:gkfsigmaeps}), Simpson-Visser (second plot on Figs.~\ref{fig:gkfsigmalamw} and \ref{fig:gkfsigmalam}), and $r_0$-model
(first plot on Figs.~\ref{fig:gkfsigmalamw} and \ref{fig:gkfsigmalam}).
Interestingly all these plots show a similar qualitative behavior.
The corrections to the graybody factor are negative at relatively low frequencies and positive for frequencies close to the maximum of the potential barrier.
These features come from the fact that, in all these cases, the deformations produce a lower though wider potential barrier
than Schwarzschild.
In summary, as compared to a Schwarzschild black hole of the same size, all these black-hole models present a more transparent
potential for high-frequency modes (a larger graybody factor, i.e., closer to one), while the barrier is more opaque and the transmission is thus suppressed
for low-frequency modes (a smaller graybody factor, i.e., closer to zero).
The $\Delta$-model \eqref{eq:deltametrik} is an exception to the commented general trend.
Its analysis is presented in App.~\ref{appA},
and it shows just the opposite behavior, inducing positive or negative modifications to the
Schwarzschild graybody factor at low or high frequencies, respectively.

Overall, our results show that the radiative properties of modified black holes contain complementary information about their geometry. The Hawking temperature probes the near-horizon structure, whereas the graybody factor encodes modifications distributed throughout the exterior spacetime.
The framework developed here therefore provides a systematic way of translating generic geometric deformations
into radiative signatures, and offers a useful basis for comparing different models of modified
black holes on the same geometrical footing.

\section*{Acknowledgments}
This work has been supported by the Basque Government Grant
\mbox{IT1977-26} and by the Grant PID2021-123226NB-I00 (funded by
MCIN/AEI/10.13039/501100011033 and by ``ERDF A way of making Europe'').

\appendix

\section{Asymptotic values of the corrections}\label{appB}

In the following, in order to obtain clear analytic results
concerning the sign of the corrections,
we will analyze in detail two interesting cases: $\ell=0$, which corresponds
to the s-waves, i.e., the modes that reflect the symmetry of the background, and large $\ell\gg\nu^2$. Let us begin with s-waves.

\paragraph{S-waves $(\ell=0)$.}

\begin{figure}
         \centering
         \begin{minipage}{0.45\linewidth}
    \includegraphics[width=\textwidth]{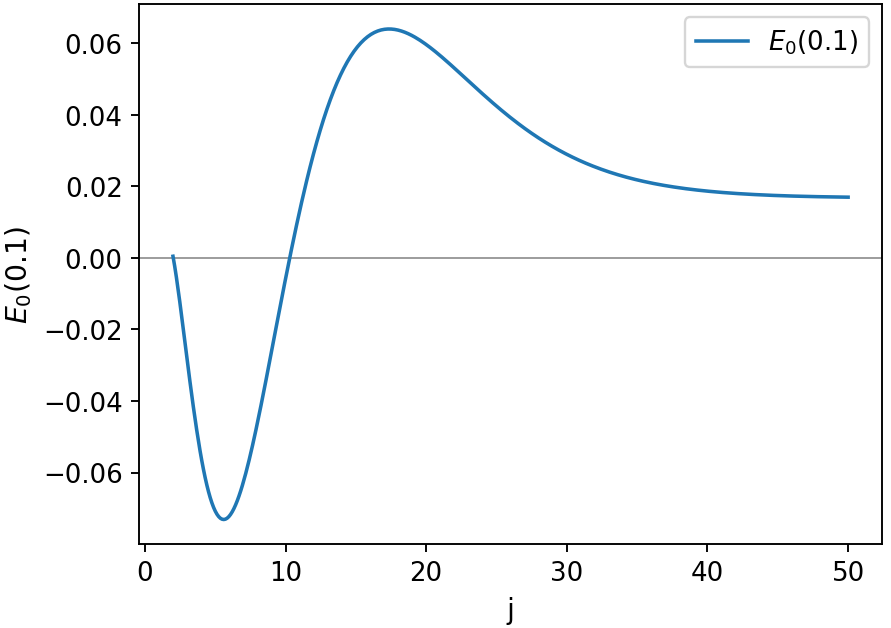}
         \caption{$E_0(0.1)$ as a function of $j$:
         $E_0(0.1)$ takes a positive value for small $j<2.022$, then in the interval $ j\in(2.022,10.299)$ it admits a negative value,
         before turning positive for $j>10.299$.
         For larger values of $j$,
         the correction remains positive, approaching the asymptotic value $E_0(0.1)\to\frac{32\sqrt2}{27}0.1^2=0.016761$
         from above.}
         \label{fig:e010}
         \end{minipage}
         \hfill
           \begin{minipage}{0.45\linewidth}
   \centering
    \includegraphics[width=\textwidth]{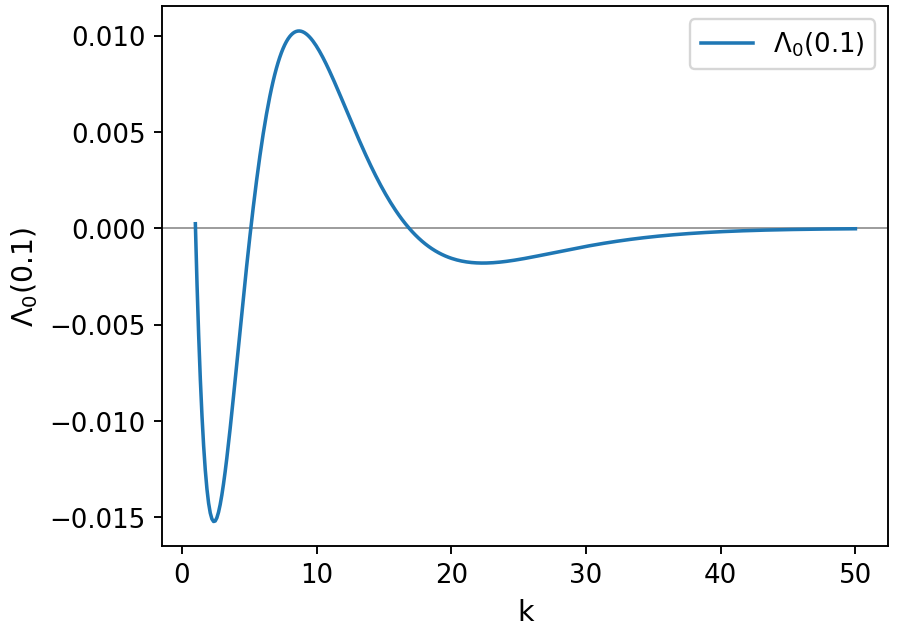}
         \caption{$\Lambda_0(0.1)$ as a function of $k$: the correction starts negative for small values of $k$ in
         the interval $k\in(1.009, 5.116)$,
         then goes to positive values in $ k\in(5.116,16.876)$, before becoming negative again.
         In contrast to $E_0(\nu)$,
         the $\lambda$-correction asymptotes to zero for large $k$, i.e., $\Lambda_0(\nu)\to0$ from below, for all $\nu\in\mathbb{R}$.}
         \label{fig:y equals x}
               \end{minipage}
     \end{figure}

For $\ell=0$ the form of the above coefficients \eqref{eq:Koln}--\eqref{eq:Lln} collapses tremendously and simplifies to
\begin{eqnarray}\label{eg:wkbfrequ}
\overset{\circ}K_0(\nu)&=&  \frac{27-256 \nu ^2}{216 \sqrt{2}},\\
    E_0(\nu)&=& -\frac{3^j \left(256 \nu ^2 (j ((j-14) j+7)+54)-27 (j-1) ((j-13) j+18)\right)-9 \nu ^2 4^{j+5}}{243\cdot2^{2
   j+\frac{9}{2}}},\label{eq:e0} \\
    \Lambda_0(\nu)&=& \frac{ 3^{k-5} \left(256 (k ((k-23) k+82)+72) \nu ^2-27 (k ((k-23) k+106)-72)\right)}{ 2^{2 k+\frac{15}{2}}}\label{eq:lam0},
\end{eqnarray}
where the individual contributions coincide with the terms presented in Ref.~\cite{Alonso-Bardaji:2025qft}.
With $\ell$ evaluated to $0$, these are just functions of the frequency $\nu$ and the corresponding power $j$ or $k$.
In Figs.~\ref{fig:e010}--\ref{fig:y equals x} we plot these quantities as functions of their corresponding
power $j$ or $k$, for a fixed frequency $\nu=0.1$.
As can be seen in these plots, the signs of $E_0(0.1)$ and $\Lambda_0(0.1)$
are not fixed and they change with $j$ and $k$, respectively. The plotted case is just an example, and
the location of these sign changes depends heavily on the specific frequency.
However, we observe that, for small values of $j$ and $k$, both $E_0$ and $\Lambda_0$ are
of the form $-a+b\nu^2$, with positive $a$ and $b$, and thus, they are negative (positive) for relatively
low (high) frequencies. While this remains true for larger $j$ in $E_0(\nu)$, things become more complicated for $\Lambda_0(\nu)$ when increasing the value of the parameter $k$. More precisely, $a$ and $b$ remain positive up to $k\approx 5$, they both turn negative for $5\lesssim k \lesssim18$, and then positive again for $k\gtrsim18$. 
At around $k\approx 18$, we find positive $a$ but negative $b$. This means that increasing the frequency in these s-waves eventually produces a sign change in $\Lambda_0(\nu)$, except for $k\approx18$, where it remains always negative.

We also find this behavior for other modes. For instance, for $\ell=1$ the limiting cases are $k\approx3$ and $k\approx14$. This last value is again the only one where $\Lambda_1(\nu)$ is negative for any $\nu$. As shown below, in the large $\ell$ limit, the zero-crossings of the $a$ and $b$ coefficients happen at $k\approx3.228$ and $k\approx11.772$. If we consider small integer values of the power $k$ (for instance, $k\leq10$), the values $k=4$ and $k=5$ are thus distinctive because $\Lambda_0(0)<0$ with ${\rm d}\Lambda_0/{\rm d}\nu>0$ and $\Lambda_{\ell\neq0}(0)>0$ with ${\rm d}\Lambda_{\ell\neq0}/{\rm d}\nu<0$ (as can be seen in the last plot in Fig.~\ref{fig:smL}). Every other value of $k\leq 10$ shows the same sign of $\Lambda_{\ell}(0)$ for any $\ell$
(as can be seen in the first three plots of Fig.~\ref{fig:smL}).

Finally, note that the asymptotic values of the corrections for large powers are $\lim_{k\to\infty}\Lambda_0(\nu)=0^-$,
which is independent of $\nu$, and $\lim_{j\to\infty}E_0(\nu)=\frac{32\sqrt2}{27}\nu^2$.

\paragraph{Large-$\ell$ limit.}

Regarding the second interesting regime, i.e., $\ell\gg\nu^2$, we compute the leading contributions to the WKB frequency according to \eqref{eq:Koln}--\eqref{eq:Lln},
\begin{eqnarray}
\overset{\circ}K_{\ell}(\nu)&=&\frac{11 \ell}{32 \sqrt{6}}+\mathcal{O}(\ell^{0})\\
  E_{\ell}(\nu)&=&-2^{j-3}3^{-j-1}(j^2-3j+2)\ell+\mathcal{O}(\ell^{-1})\label{eq:eun}\\
   \Lambda_{\ell}(\nu)&=&-\frac{2^{k-4}  k \left(k^2-15 k+38\right)}{3^{k+3}\ell}+\mathcal{O}(\ell^{-3}),\label{eq:lamun}
\end{eqnarray}
which turn out to be independent of the frequency $\nu$.
By comparing directly their magnitude, one finds that the $\Lambda_\ell(\nu)$ contribution is several orders of magnitude smaller than $E_\ell(\nu)$. In fact, $E_{\ell}(\nu)$ seems to be always dominant, especially for large $\ell$ because $|E_\ell(\nu)|\sim\ell$ while $|\Lambda_{\ell}(\nu)|\sim\frac1\ell$.
This implies that, as already commented several times, presuming $\frac\lambda\epsilon=\mathcal{O}(1)$,
the correction $\delta\sigma^\lambda_{\ell}(\nu)$ is naturally smaller than $\delta\sigma^\epsilon_{\ell}(\nu)$ in magnitude.
It should be noted that our statements are tied to the assumption that $\nu$ is such that \eqref{eq:gkb-wkb} is real, that is, that the squared frequency is smaller than the maximum of the potential barrier.

Concerning the signs of the correction in this large-$\ell$ limit, it turns out that
the sign of $E_\ell(\nu)$ is completely fixed for $j>2$ and any frequency $\nu$, that is,
$E_\ell(\nu)<0$. However, $\Lambda_\ell(\nu)$ changes sign depending on $k$:
it is positive in the interval $3.228\lesssim  k\lesssim  11.772$, while, for small and large
$k$, that is $k\lesssim 3.228$ and $k\gtrsim11.772$, $\Lambda_\ell(\nu) <0$.

\section{Example model beyond the simple Schwarzschild deformation}\label{appA}

Let us briefly analyze the $\Delta$-model \eqref{eq:deltametrik}, which
does not fit the simple ansatz considered in the main part of the article for slightly deformed
Schwarzschild geometries.
Note that, for this model, the shape functions $h_1(r)=1$ and $h_2(r)=r$ coincide with their Schwarzschild
values, while
\begin{equation}
h_3(r)=1+\frac{\Delta}{r^2}\left( 1- \frac{2 M}{r}\right),
\end{equation}
which, in terms of the dimensionless radius $x=r/(2M)$, reads
\begin{equation}
h_3(x)=1+\frac{\Delta}{(2 M)^2x^2}\left( 1- \frac{1}{x}\right).
\end{equation}
Therefore, this can not be described by the ansatz \eqref{defh3}, as it contains two terms with
different powers $k=2$ and $k=3$. At this point, it is thus unclear whether one of the terms
dominates around the maximum of the potential and the other one may be discarded in the computation
of the graybody factor under the approximation performed in the paper, or both contribute and none of
them can be discarded.

Hence,
in the following, we will perform the complete analytic derivation of the graybody factor,
for which our analysis in Sec.~\ref{sec.graybody} remains applicable, though
we just have to consider the full form of $h_3(r)$.
However, we assume the parameter $\bar{\lambda}:=-\Delta/(2M)^2$
to be small,
and expand the graybody factor of this model as a small deformation around the Schwarzschild graybody factor---just as in Eq.~\eqref{eq:gkf01}---,
\begin{equation}
    \sigma_\ell(\nu)=\overset{\circ}\sigma_\ell(\nu)
+2\pi\overset{\circ}\sigma_\ell(\nu)\left(1-\overset{\circ}\sigma_\ell(\nu)\right)\bar{\lambda}\,\bar{\Lambda}_\ell(\nu),
\end{equation}
where the correction term is given by the lengthy expression
\begin{align}
\bar{\Lambda}_\ell(\nu)
  &= -\frac{8\sqrt2}
      {\big(\ell(\ell+1)\big)^{3/2}
       \big(3\ell(\ell+1)+W-3\big)^{4}
       \big(9-3W+\ell(\ell+1)(14+9\ell(\ell+1)+3W)\big)^{5/2}}
      \notag\\[4pt]
  &\quad\times\Big[1620\,\ell^{20}+16200\,\ell^{19}+\ell^{18}\,\Big(540 W + 73596 -19683\,\nu^2\Big) +\ell^{17}\,\Big(4860 W + 200664 -177147\,\nu^2\Big) \notag\\
&\quad +\ell^{16}\,\Big(19252 W + 361236 - (6561 W + 707454)\,\nu^2\Big) +\ell^{15}\,\Big(43856 W + 434544 - (52488 W + 1644300)\,\nu^2\Big) \notag\\
&\quad +\ell^{14}\,\Big(64320 W + 299480 - (178227 W + 2455614)\,\nu^2\Big)\notag\\
&\quad +\ell^{13}\,\Big(68320 W - 54016 - (329049 W + 2467710)\,\nu^2\Big) \\
&\quad +\ell^{12}\,\Big(68776 W - 506456 - (349572 W + 1720556)\,\nu^2\Big) +\ell^{11}\,\Big(87864 W - 879488 - (207999 W + 859440)\,\nu^2\Big) \notag\\
&\quad +\ell^{10}\,\Big(120104 W - 1028648 - (68505 W + 303417)\,\nu^2\Big) +\ell^{9}\,\Big(135440 W - 920080 - (23382 W + 55009)\,\nu^2\Big) \notag\\
&\quad +\ell^{8}\,\Big(116832 W - 646348 - (7910 W + 56910)\,\nu^2\Big) +\ell^{7}\,\Big(76544 W - 354664 + (5989 W - 233616)\,\nu^2\Big) \notag\\
&\quad +\ell^{6}\,\Big(37404 W - 147140 + (19755 W - 597278)\,\nu^2\Big) +\ell^{5}\,\Big(12396 W - 41976 + (52906 W - 997326)\,\nu^2\Big) \notag\\
&\quad +\ell^{4}\,\Big(2052 W - 6156 + (128743 W - 1275528)\,\nu^2\Big)+\ell^{3}\,\Big((163593 W - 1143180)\,\nu^2\Big) \notag\\
&\quad +\ell^{2}\,\Big((158598 W - 793395)\,\nu^2\Big)+\ell\,\Big((86751 W - 352107)\,\nu^2\Big) +\Big((39366 W - 118098)\,\nu^2\Big)\notag
      \Big],
\end{align}
in which we have defined the shorthand notation
\begin{equation}
    W := \sqrt{\,9+\ell(\ell+1)\big(14+9\,\ell(\ell+1)\big)\,}\, .
\end{equation}
Albeit hideous, this expression can be studied in the same way as the previous expressions
in the main body of the paper. We are particularly interested in the sign of the correction term here.
For the sake of completeness, we give the s-wave expression,
\begin{equation}
    \bar\Lambda_0(\nu)=\sqrt2\left(\frac{97}{864} \nu^2-\frac{13}{8192}\right),
\end{equation}
and the large $\ell$-limit,
\begin{equation}
    \bar\Lambda_\ell(\nu)=-\frac{5}{243\ell}+\mathcal{O}(\ell^{-2}).
\end{equation}
For s-waves, it becomes clear that $\bar\Lambda_0(\nu)$ is positive (negative) for large (small) enough values of $\nu$.
However, for large modes $\ell\gg\nu^2$, the correction turns out to be negative, independently of the frequency.

\begin{figure}
    \centering
\includegraphics[width=0.5\linewidth]{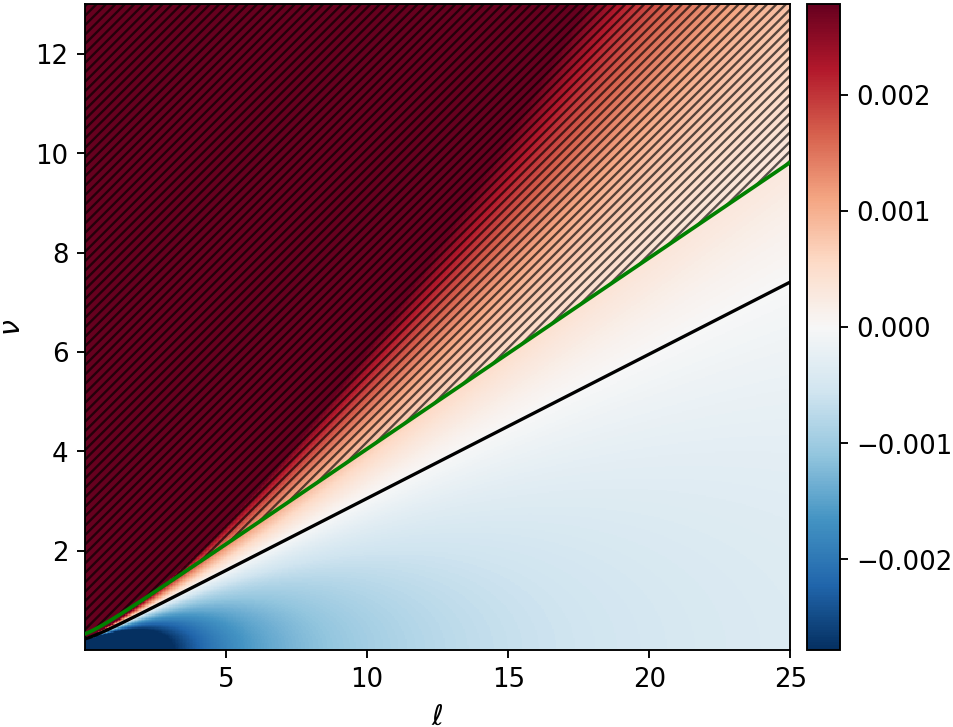}
    \caption{
    Sign map for the correction term $\bar\Lambda_\ell(\nu)$ over the ranges $\ell\in[0,25]$ and $\nu\in[0,13]$. Red (blue) regions indicate $\bar\Lambda_\ell(\nu)>0$ ($\bar\Lambda_\ell(\nu)<0$), separated by the solid black curve marking the exact zero-crossing $\bar\Lambda_\ell(\nu)=0$. The green curve shows the boundary $\nu^2=\mathring v_\ell(\mathring x_{\max})$, above which the zeroth-order (Schwarzschild) quantity $\mathring d_{\max}^\ell(\nu)$ ceases to be real; the hatched region marks where this occurs.}
    \label{fig:figure15}
\end{figure}

In particular, Fig.~\ref{fig:figure15} shows that the sign map of $\bar\Lambda$ resembles the one in Fig.~\ref{fig:smL} for the value $k=3$. This makes sense from the perspective of $ h_3(x)=1-\frac{\bar{\lambda}}{x^3}\left( x- 1\right)$, 
because between the turning points, near the maximum of the potential $x\approx x_{\rm max}$, which always lies in the interval
$4/3\leq x_{\max}\leq3/2$, the term between parenthesis is a positive quantity of order one
that varies within a relative narrow range, and thus
\begin{equation}
 h_3(x)\approx1-\frac{c\bar{\lambda}}{x^3},
\end{equation}
with $c\approx1$.
Therefore, if one wishes to understand this model in terms of the simple ansatz \eqref{h3intermsoflambda}, from this last expression
one could read off the (approximate) parameters $\lambda\approx c\bar\lambda=-c\Delta/r_H^2<0$ and $k\approx 3$.

In any case,
since the correction parameter $\bar\lambda$ is negative definite,
the modification to the Schwarzschild graybody factor has the opposite sign to that of $\bar{\Lambda}_\ell(\nu)$.
In conclusion, this model shows just the opposite behavior as compared to the other models analyzed in the paper.
More precisely, for large frequencies (close to the maximum of the potential) the graybody factor is smaller
than its Schwarzschild counterpart,
while for small frequencies it is larger. Hence, the transmission coefficient of the outgoing high-frequency
(low-frequency) modes is diminished (enhanced).

\bibliographystyle{bib-style}
\bibliography{biblio}

\end{document}